\documentclass[aps,prd,reprint,nofootinbib]{revtex4-2}

\usepackage{subcaption}
\usepackage{graphicx}% Include figure files
\usepackage{dcolumn}% Align table columns on decimal point
\usepackage{bm}% bold math
\usepackage{amsmath}
\usepackage{multirow}
\usepackage[dvipsnames]{xcolor}
\usepackage{natbib}
\usepackage{ragged2e}
\usepackage{url}
\usepackage{caption}

\usepackage[colorlinks=true, linkcolor=blue, urlcolor=blue, citecolor=blue]{hyperref}

\usepackage{booktabs}
 
\begin{document}

\preprint{PRD}

\title{Energy Deposition by Galactic Cosmic Rays and Implications for Ozone Chemistry}

%\title{Monte Carlo Simulation of Galactic Cosmic Ray Energy Deposition and Stratospheric Chemistry}

%\title{Monte Carlo Simulation of Galactic Cosmic Ray Cascades in Earth's Stratosphere: Particle Production and Energy Deposition}

%\title{\texttt{Geant4} Modeling of Galactic Cosmic Ray Cascades in the Stratosphere and Their Impact on Ozone Chemistry}% Force line breaks with \\

\author{Luiz A. Stuani Pereira$^{1,2}$}
\email{luizstuani@uaf.ufcg.edu.br}

\author{Rita C. Anjos$^{3,4,5,6,7}$}%
\email{ritacassia@ufpr.br}

\affiliation{$^1$Instituto de Física da Universidade de São Paulo (IFUSP), \& R. do Matão, 1371, São Paulo, 05508-090, SP, Brazil.}
\affiliation{$^12$Unidade Acadêmica de Física da Universidade Federal de Campina Grande (UAF-UFCG), \& R. Aprígio Veloso, 882, Campina Grande, 58429-900, PB, Brazil.}

\affiliation{$^3$Departamento de Engenharias e Exatas, Universidade Federal do Paraná (UFPR), Rua Pioneiro, Palotina, 85950-000, PR, Brazil.}
\affiliation{$^4$Programa de pós-graduação em Física \& Departamento de Física, Universidade Estadual de Londrina (UEL), Rodovia Celso Garcia Cid Km 380, Londrina, 86057-970, PR, Brazil.}
\affiliation{$^5$Programa de Pós-Graduação em Física e Astronomia, Universidade Tecnológica Federal do Paraná (UTFPR), Av. Sete de Setembro, 3165, Curitiba, 80230-901, PR, Brazil.}
\affiliation{$^6$Programa de Pós-Graduação em Física Aplicada, Universidade Federal da Integração Latino-Americana, Av. Tarquínio Joslin dos Santos, 1000, Foz do Iguaçu, 85867-670, PR, Brazil.}
\affiliation{$^7$Departamento de Física, Universidade Federal do Espírito Santo, Núcleo de Astrofísica e Cosmologia (Cosmo-Ufes), Av. Fernando Ferrari, 514, Vitória, 29075-910, ES, Brazil.}


\date{\today}% It is always \today, today,

\begin{abstract}

We present a Monte Carlo study of galactic cosmic-ray (GCR) energy deposition and its implications for stratospheric chemistry, performed with the \texttt{Geant4} toolkit. Primary nuclei (protons, $\alpha$, CNO, and Si) were propagated through an atmosphere modeled from 0 to 120~g~cm$^{-2}$, considering both Polar ($R_{\mathrm{c}}=0.1$~GV) and Equatorial ($R_{\mathrm{c}}=15$~GV) geomagnetic cutoff conditions. The simulations resolve the variation of energy deposition with altitude for primary and secondary particles, revealing that $\sim$~96\% of the stratospheric energy budget arises from cascade secondaries within the 15--35~km domain. By converting layer-resolved energy deposition into ion pair production rates, we quantify the resulting formation of odd nitrogen (NO$_{\rm x}$) and odd hydrogen (HO$_{\rm x}$) radicals, which catalyze the destruction of ozone. The modeled production rates peak between 18 and 22~km altitude, leading to an estimated fractional ozone decrease of order $10^{-3}$--$10^{-2}$ under average GCR fluxes, consistent with observed background modulation over the solar cycle. These results establish a physically consistent link between cosmic-ray induced energy deposition and ozone chemistry, providing a benchmark framework for coupling high-energy particle transport to atmospheric photochemical models.

\end{abstract}


\maketitle

%\tableofcontents

\section{\label{sec:introd}Introduction}

Galactic cosmic rays (GCRs) constitute a continuous flux of high-energy particles, primarily protons, alpha particles, and a spectrum of medium to heavy nuclei produced in astrophysical sources such as supernova remnants~\cite{2019IJMPD..2830022G}. Upon entering Earth's atmosphere, these primaries interact with atmospheric nuclei, initiating nucleonic-electromagnetic cascades that produce secondary neutrons, pions, muons, electrons, and photons. These interactions deposit energy and generate ionization throughout the atmospheric column~\cite{Dorman2004, Usoskin2006}. The resulting ion pair production influences multiple atmospheric processes, including the formation of odd nitrogen (NO$_{\rm x}$) and odd hydrogen (HO$_{\rm x}$), which are key agents in catalytic ozone destruction cycles~\cite{Porter1976, Vitt1996}. Because the stratospheric ozone layer both shapes the radiative energy balance and shields the biosphere against harmful ultraviolet radiation, quantifying cosmic-ray-induced energy deposition and subsequent ionization is fundamental to understanding its role in ozone modulation.

\begin{figure}[tb]
 \centering
 \includegraphics[width=0.5\textwidth]{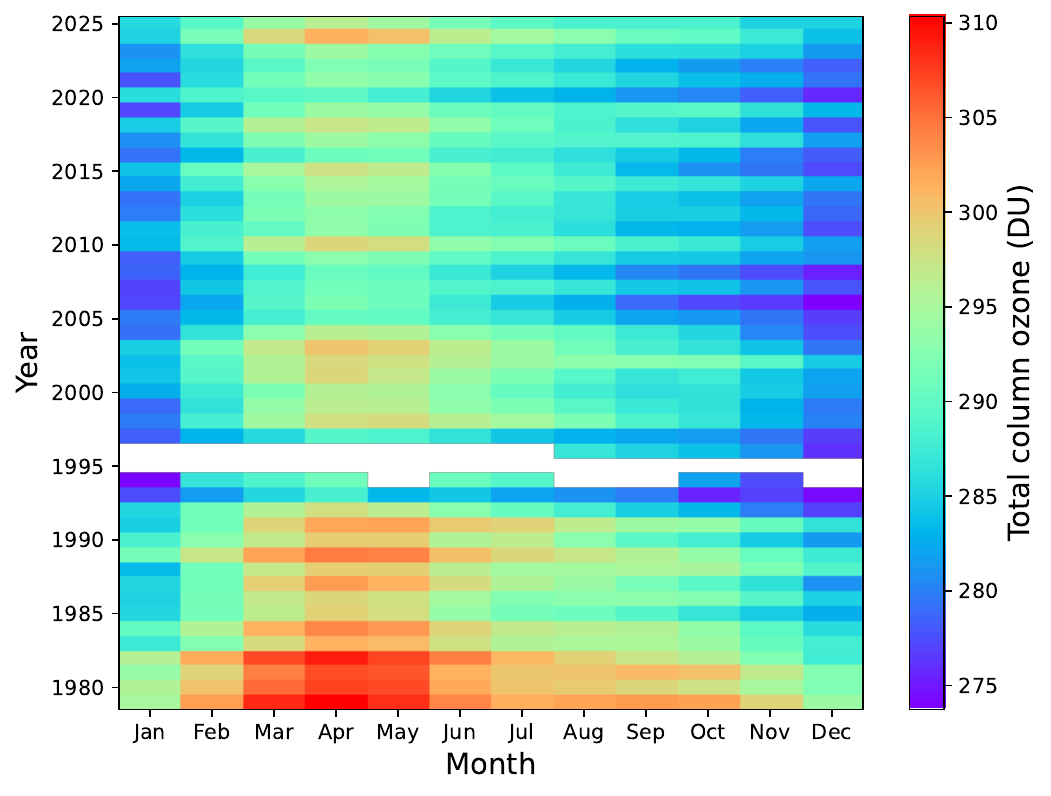}
 \caption{\small \justifying Monthly mean total column ozone (in Dobson Units, DU) as a function of latitude ($90^{\circ}$S--$90^{\circ}$N) and time (1979--2025), derived from the combined TOMS, OMI, and OMPS satellite records. The color scale indicates the ozone column density, with warmer colors corresponding to higher values. White cells denote missing data.}
\label{fig:ozone_mean}
\end{figure}

The long-term global monthly mean total column ozone (Figure~\ref{fig:ozone_mean}) reveals a distinct annual cycle, with ozone maxima (warmer colors, 300--315 DU) typically occurring during the boreal and austral spring months, and minima (cooler colors, 275--290 DU) appearing in autumn. This climatological pattern is modulated by stratospheric dynamics, chemical processes, and external forcings. Superimposed on the annual cycle is a pronounced multi-decadal trend: ozone values were generally lower in the 1990s and early 2000s, reflecting the period of peak anthropogenic ozone depletion, followed by a gradual recovery in recent decades after the implementation of the Montreal Protocol. Regional and year-to-year variability, such as the anomalously high ozone columns in the Northern Hemisphere spring of 2020, is also evident and can be linked to dynamical disturbances in the stratosphere. The white gap around 1994--1996 corresponds to years with missing satellite retrievals in the combined Total Ozone Mapping Spectrometer (TOMS), Ozone Monitoring Instrument (OMI), and Ozone Mapping and Profiler Suite (OMPS) record~\cite{nasa_ozone}. This record provides the essential observational context against which the more subtle potential influences of cosmic-ray-induced ionization on ozone variability can be investigated.

The ionization rate driven by GCRs peaks near the Pfotzer maximum ($\sim$15--18 km altitude), coinciding with the lower-stratospheric ozone concentration maximum~\cite{Calisto2011, 2014HGSS....5..175C}. This spatial concurrence has motivated numerous studies exploring potential links between cosmic-ray variability and ozone depletion~\cite{Calisto2011, Jackman2016}. Models such as \texttt{CORSIKA} (COsmic Ray SImulations for KAscade) and \texttt{CRAC:CRII} (Cosmic Ray Atmospheric Cascade: Cosmic Ray Ionization) have simulated these cascades in detail, providing the basis for ionization yield functions across atmospheric depths~\cite{Usoskin2006}. Nonetheless, the precise quantification of energy deposition, especially from heavier nuclei such as nitrogen and silicon, remains an open question. Their larger mass and charge produce distinctive energy-loss per path-length profiles ($dE/dx$) that can modify the local ionization structure, possibly influencing ozone chemistry via altered NO$_{\rm x}$--HO$_{\rm x}$ production efficiencies.

Recent three-dimensional chemistry--climate model (CCM) studies, using ionization parameterizations like \texttt{CRAC:CRII}, have demonstrated that cosmic rays can induce NO$_{\rm x}$ enhancements of up to 10\% in the tropopause region and concomitant ozone reductions of 1--3\% in the lower stratosphere~\cite{Calisto2011}. Jackman et al.\ (2016)~\cite{Jackman2016} confirmed similar findings using the SD-WACCM and GSFC 2-D models, attributing ozone losses of 0.2--1\% primarily to GCR-induced NO$_{\rm x}$ increases (1--6\%) between 100--10~hPa. These simulations also noted that the magnitude of GCR effects depends strongly on solar modulation, aerosol loading, and stratospheric chlorine concentration. During solar minima, when GCR flux is greatest, the reduced heliospheric magnetic shielding amplifies ionization rates and the corresponding chemical perturbations. Even though the overall global-mean ozone impact remains modest ($<$1\%), these variations represent a persistent natural forcing on stratospheric chemistry that must be quantified and understood across solar cycles.

An alternative perspective has emerged from laboratory and modeling work by Lu et al.~\cite{Lu2001, Lu2009, Lu2010, Lu2015}, who proposed the cosmic-ray-driven electron-induced reaction (CRE) mechanism. In this view, low-energy secondary electrons produced by GCR cascades interact with halogenated molecules (e.g., CFCs) adsorbed on the surfaces of polar stratospheric clouds (PSCs), triggering dissociative electron-transfer reactions that release reactive chlorine atoms. These catalytic Cl-based reactions can rapidly destroy O$_3$, particularly in the cold, dense Antarctic vortex environment. The CRE model successfully reproduced the observed 11-year cyclic variations of Antarctic ozone loss and stratospheric temperature minima, implying that cosmic-ray modulation might play a greater role in ozone variability than photochemical models alone predict~\cite{Lu2015}. However, this mechanism remains under debate. Subsequent observational studies have reported pronounced 11-year cyclic variations in Antarctic ozone loss and lower-stratospheric temperatures over four solar cycles \cite{Lu2021}, and a parameter-free quantitative formulation of the CRE theory has since been developed \cite{Lu2023, Lu2025}. Nevertheless, physically grounded models quantifying particle energy deposition remain essential for independently assessing the magnitude of cosmic-ray forcing on ozone chemistry.

%However, this mechanism remains under debate, given the contrasting results from empirical analyses~\cite{Patra2002} that highlight the need for physically grounded models to quantify particle energy deposition, rather than relying solely on correlations.

Three decades of satellite composites and regression analyzes reveal ozone dynamics of considerable complexity that cannot be attributed solely to declining halogen loading. While upper-stratospheric ozone ($>$10~hPa) has exhibited statistically significant recovery since the late 1990s following the implementation of the Montreal Protocol, the lower stratosphere continues to show a persistent negative trend at mid-latitudes (60$^{\circ}$S--60$^{\circ}$N)~\cite{Bourassa2014, Ball2018}. These downward trends are particularly significant because they occur precisely in the altitude band where GCR-induced ionization peaks, raising fundamental questions about the role of high-energy particle interactions in driving ongoing ozone variability. This is especially relevant in the context of long-term stratospheric cooling and structural changes in the Brewer-Dobson circulation, both of which modulate the transport and chemical lifetime of ozone-depleting species.

To advance this understanding, it is necessary to compute detailed energy deposition profiles from realistic cosmic-ray compositions and spectra, and to map them layer by layer through the atmosphere. Previous Monte Carlo studies have demonstrated that ionization and secondary-particle production vary sharply with both primary particle type and energy~\cite{Usoskin2006}. While analytical and semi-empirical models often approximate cosmic-ray fluxes using proton-dominant compositions, heavier nuclei like helium, nitrogen, and silicon contribute disproportionately to energy deposition because their stopping power ($dE/dx$) scales approximately with $Z^2/A$, where $Z$ is the atomic number and $A$ is the mass number. Consequently, the precise determination of ionization yields from these components is crucial for constraining the local energy input available for triggering ozone-affecting chemical reactions.

In this study, we developed a dedicated \texttt{Geant4}-based model of the stratosphere~\cite{Agostinelli2002, Desorgher2005} specifically designed to quantify the impact of GCR-induced ionization on ozone depletion. Unlike general atmospheric transport simulations, our model focuses exclusively on the critical 15--35 km altitude domain, resolving the microphysical interactions of primary GCR constituents (protons ($^1$H), alpha particles ($^4$He), CNO-group nuclei, and silicon-group nuclei) as they traverse the ozone-rich layers. A central objective of this work is to evaluate the stratospheric response under two distinct geomagnetic conditions: a \textit{Polar scenario} ($R_{\rm c} \approx 0.1$ GV), where the atmosphere is exposed to the full, solar-modulated GCR spectrum, and an \textit{Equatorial scenario} ($R_{\rm c} \approx 15$ GV), where the geomagnetic field effectively shields the atmosphere from low-rigidity particles.

The core utility of the present model lies in its capacity to establish a quantitative bridge between microphysical particle transport and macroscopic atmospheric chemical consequences. The simulated layer-resolved energy deposition profiles are converted into altitude-dependent ion pair production rates, which constitute the primary source term for the formation of odd nitrogen (\(\mathrm{NO_x} = \mathrm{NO} + \mathrm{NO_2}\)) and odd hydrogen (\(\mathrm{HO_x} = \mathrm{OH} + \mathrm{HO_2}\)) radicals. These species are the dominant catalysts in the well-established odd-nitrogen and odd-hydrogen ozone destruction cycles. By resolving the energy input on a layer-by-layer basis, rather than relying on column-integrated flux approximations, this approach provides geometry-specific ionization datasets suitable for direct coupling to photochemical and chemistry-climate models, thereby advancing the physical grounding of GCR--ozone interaction studies.

This paper is organized as follows. Section~\ref{sec:methodology} details the construction of the \texttt{Geant4} stratospheric model~\citep{Agostinelli2002}, including the spherical geometry, the density profile, and the dry-air composition implemented across four concentric layers (15--35~km). It also describes the parameterization of the primary GCR spectra for both Polar and Equatorial rigidities and the physics models employed for hadronic and electromagnetic interactions. Section~\ref{sec:results} presents the simulation results, beginning with a comparative analysis of the total and layer-resolved energy deposition. This is followed by an examination of the volumetric production of secondary particles and the emerging particle flux. The altitude-dependent ionization energy flux is then converted into ion pair production rates and subsequently into the chemical formation of odd nitrogen (NO$_{\rm x}$) and odd hydrogen (HO$_{\rm x}$) radicals. From these quantities, we estimate the corresponding ozone perturbation using established photochemical parameterizations, comparing our results with climate models such as SD-WACCM (Specified Dynamics Whole Atmosphere Community Climate Model) and GSFC 2-D (Goddard Space Flight Center Two-Dimensional Coupled Chemistry-Dynamics-Radiation Model)~\cite{Jackman2016}. Finally, Section~\ref{sec:summary} summarizes the main findings and discusses the broader implications of using dedicated Monte Carlo transport models to quantify stratospheric composition changes under varying geomagnetic shielding conditions.

\section{Methodology}
\label{sec:methodology}

The simulations were carried out using the \texttt{Geant4} Monte Carlo toolkit (version 11.3.2) to model the transport and interaction of GCR nuclei in a stratified atmospheric geometry representative of the ozone-bearing stratosphere~\citep{Agostinelli2002, Desorgher2005, Allison2006}. We designed the model to achieve two aims: to resolve the depth-dependent energy deposition and secondary-particle production resulting from GCR primaries within the 15-35 km altitude range, and to relate the deposited energy flux to the potential for ozone perturbation via ion-chemical pathways. The simulations combine physically realistic geometry and composition data, isotropic primary-particle injection following observed energy spectra, and standard heliospheric and geomagnetic modulation corrections. Each simulation output provides the energy deposited and multiplicity distributions per atmospheric layer, which are subsequently analyzed to infer ion pair production rates and relative ionization yields. The following subsections provide a detailed description of the atmospheric modeling, primary-particle specification, and spectral modulation schemes employed in this study.

\subsection{Atmospheric Geometry and Composition}
\label{sec:atmosphere}

The atmospheric geometry was modeled as four concentric spherical shells representing stratified layers between 15 and 35 km in altitude, a region encompassing the lower and middle stratosphere, where both ozone concentration and cosmic-ray-induced ionization are significant. The layers were centered at the origin of an Earth-centered coordinate system. This configuration simplifies the injection of isotropic primaries from a bounding spherical surface, ensuring numerical stability in the ray-tracing routines.

\begin{figure}[tb]
 \centering
 \includegraphics[width=0.5\textwidth]{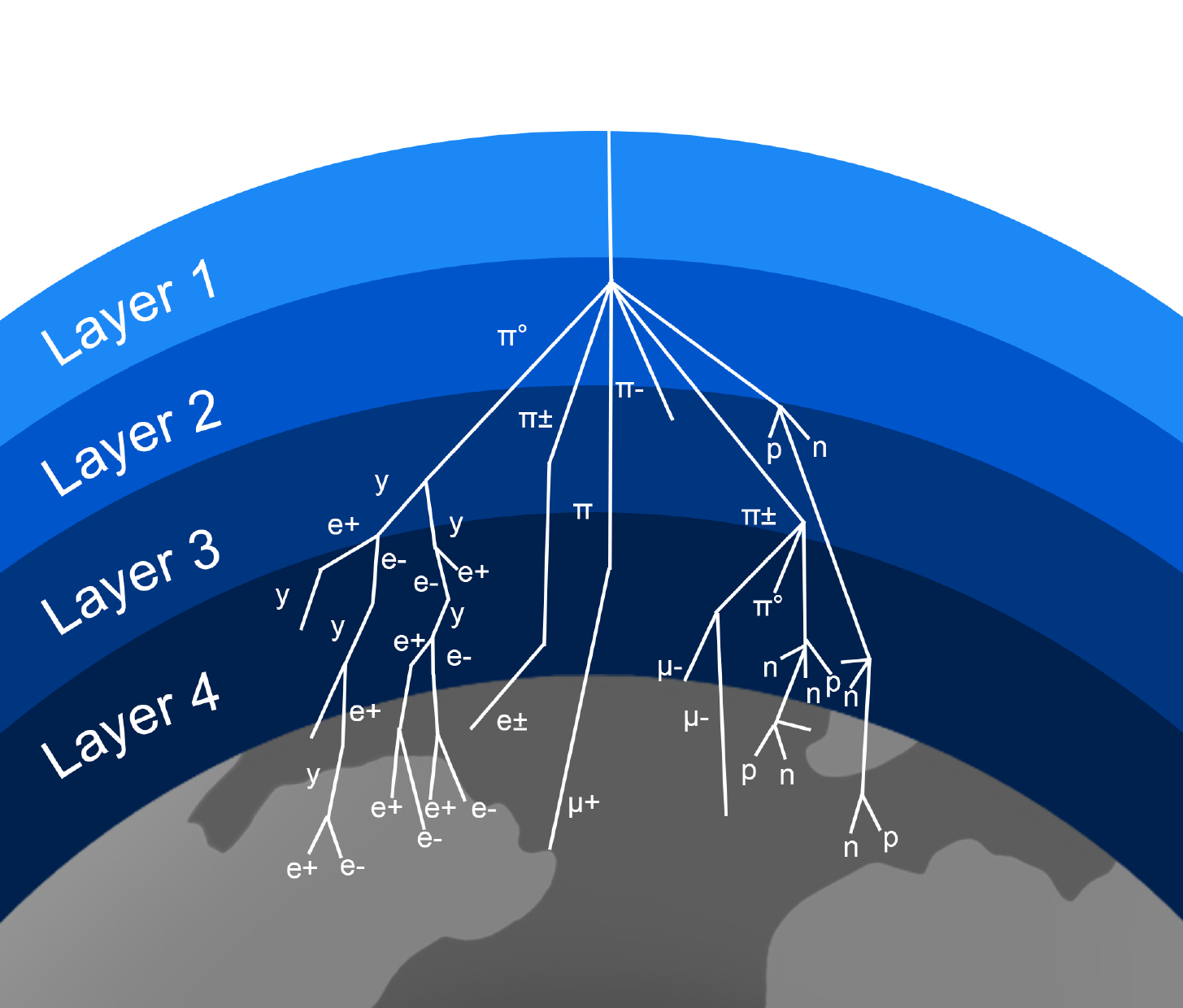}
 \caption{\small \justifying Two-dimensional wireframe representation of the atmospheric geometry implemented in the \texttt{Geant4} simulations. The model consists of four concentric spherical layers corresponding to the altitude ranges 15--20 km, 20--25 km, 25--30 km, and 30--35 km, surrounding the Earth's surface.}
\label{fig1}
\end{figure}

Table~\ref{tab:layers} lists the altitude ranges, radii, and densities of each atmospheric layer. All layers share a common dry-air composition consisting of nitrogen (N), oxygen (O), argon (Ar), and a trace amount of carbon (C), with mass fractions held constant across altitudes. The density decreases exponentially with altitude, following representative values for the stratosphere under standard atmospheric conditions~\cite{USSA1976, Picone2002}. The resulting geometry is illustrated in Figure~\ref{fig1}, showing the four concentric spherical layers implemented in the \texttt{Geant4} environment.

\begin{table}[tb]
\centering
\setlength{\tabcolsep}{12pt}
\renewcommand{\arraystretch}{1.3}
\begin{tabular}{ccc}
\hline\hline
\textbf{Layer} & \textbf{Altitude (km)} & \textbf{Density (g~cm$^{-3}$)} \\
\hline
4 & 15--20 & $1.4\times10^{-4}$ \\
3 & 20--25 & $6.4\times10^{-5}$ \\
2 & 25--30 & $2.9\times10^{-5}$ \\
1 & 30--35 & $1.3\times10^{-5}$ \\
\hline\hline
\end{tabular}
\caption{\small \justifying Atmospheric layers implemented in the simulation geometry. Layer numbering increases from the outermost (Layer 1) to the innermost (Layer 4)~\cite{USSA1976, PDAS_USSA1976_Tables}. Each layer is defined as a dry-air mixture with constant elemental mass fractions $X_\mathrm{N} = 0.755$, $X_\mathrm{O} = 0.232$, $X_\mathrm{Ar} = 0.012$, and $X_\mathrm{C} = 0.0003$ across all altitudes. The densities correspond to the mean of each altitude interval as used in the \texttt{Geant4} configuration.}
\label{tab:layers}
\end{table}

This layered spherical schema reproduces the large-scale stratification necessary for cosmic-ray shower initiation and energy loss through the stratosphere. The chosen altitude range corresponds to the Pfotzer-maximum region~\cite{Calisto2011, 2014HGSS....5..175C}, where the first nuclear interactions and peak ionization occur for GeV--TeV primaries. For more comprehensive atmospheric studies, the model could be extended with additional shells or continuous density profiles based on the U.S. Standard Atmosphere~\cite {USSA1976}.

\subsection{Primary Cosmic-Ray Composition and Spectral Modeling}
\label{sec:primaries}

The incoming GCR are represented by four dominant nuclear groups: protons (p), helium nuclei (He), light nuclei (CNO group), and intermediate-mass nuclei (Si). These components collectively account for more than 99\% of the total GCR flux below the knee energy region ($\sim10^{15}$ eV) and are therefore sufficient to describe the primary population relevant for stratospheric interactions~\cite{Gaisser2016, Navas2024}.

A total of $10^4$ primary particles were simulated for each scenario, distributed according to their typical relative abundances under near-Earth GCR conditions~\citep{1993ApJ...405..567L, Calisto2011, Gaisser2016}. The composition was set as follows: protons (91\%), helium nuclei (8\%), light intermediate nuclei (CNO group, 0.7\%), and intermediate-mass nuclei (Si group, 0.3\%). Particles were injected isotropically from a spherical surface of radius $R_{\mathrm{inj}} = 36$ km (the top of the model atmosphere) directly above the outermost shell described in Section~\ref{sec:atmosphere}. The primary directions were sampled over the inward hemisphere, with polar angles $\theta \in [0^\circ, 90^\circ]$ relative to the local inward radial direction and azimuthal angles $\phi \in [0^\circ, 360^\circ]$, corresponding to a uniform distribution in solid angle toward the center of the Earth.

\subsubsection{Geomagnetic cutoff and latitudinal dependence}

To investigate the dependence of the atmospheric cascade and subsequent ionization on geographic latitude, the transmission of primary particles was modeled using a geomagnetic cutoff rigidity based on the St\"ormer approximation~\cite{Stormer1956}. In the dipole approximation, the cutoff rigidity depends on the geomagnetic latitude~$\lambda$:
\begin{equation}
  R_{\rm c}(\lambda) \simeq 14.9\,\cos^{4}\lambda \;\; \text{GV}.
  \label{eq:cutoff_rigidity}
\end{equation}
Two distinct geomagnetic scenarios were simulated to bound the range of atmospheric effects:
\begin{enumerate}
    \item \textbf{Polar Region ($\lambda \approx 90^\circ$):} Characterized by a cutoff rigidity $R_{\rm c} \approx 0.1$ GV \cite{Smart2008}. In this scenario, the geomagnetic field lines are open, allowing the full interstellar spectrum essentially (down to the solar modulation limit) to enter the atmosphere. This represents the maximum ionization potential, relevant for phenomena such as the ozone hole~\cite{Solomon1999}.
    \item \textbf{Equatorial Region ($\lambda \approx 0^\circ$):} Characterized by a high cutoff rigidity $R_{\rm c} \approx 15$ GV~\cite{Smart2008}. In this scenario, the strong perpendicular component of the geomagnetic field filters out all particles with rigidity below $\sim$15 GV (corresponding to kinetic energies below $\sim$14.1~GeV for protons).
\end{enumerate}

The geomagnetic transmission function was implemented using a smooth parameterization that accounts for the finite width of the penumbra region, where the transmission varies continuously from zero to unity~\cite{Smart2008, Desorgher2005}. Following empirical trajectory calculations, the transmission function is given by:
\begin{equation}
T_{\text{smooth}}(R, R_{\rm c}, \Delta R) = 
\begin{cases}
0, & R < R_{\rm c} - \Delta R/2, \\[4pt]
T_{\text{pen}}(R), & |R - R_{\rm c}| \leq \Delta R/2, \\[4pt]
1, & R > R_{\rm c} + \Delta R/2,
\end{cases}
\label{eq:smooth_transmission}
\end{equation}
where the penumbra transition is
\begin{equation}
T_{\text{pen}}(R) = \frac{1}{2}\left[1 + \sin\left(\frac{\pi(R - R_{\rm c})}{\Delta R}\right)\right].
\label{eq:penumbra_transition}
\end{equation}
where $R$ is the particle rigidity, $R_{\rm c}$ is the cutoff rigidity, and $\Delta R$ represents the penumbra width. Based on worldwide cutoff rigidity surveys~\cite{Smart2008}, we adopted penumbra widths of $\Delta R = 1.5$ GV for the Polar scenario and $\Delta R = 2.0$ GV for the Equatorial scenario. This smooth transition function replaces the idealized step function, providing a more realistic description of geomagnetic filtering at the top of the atmosphere.

\begin{figure*}[tb]
 \centering
\includegraphics[width=1.0\textwidth]{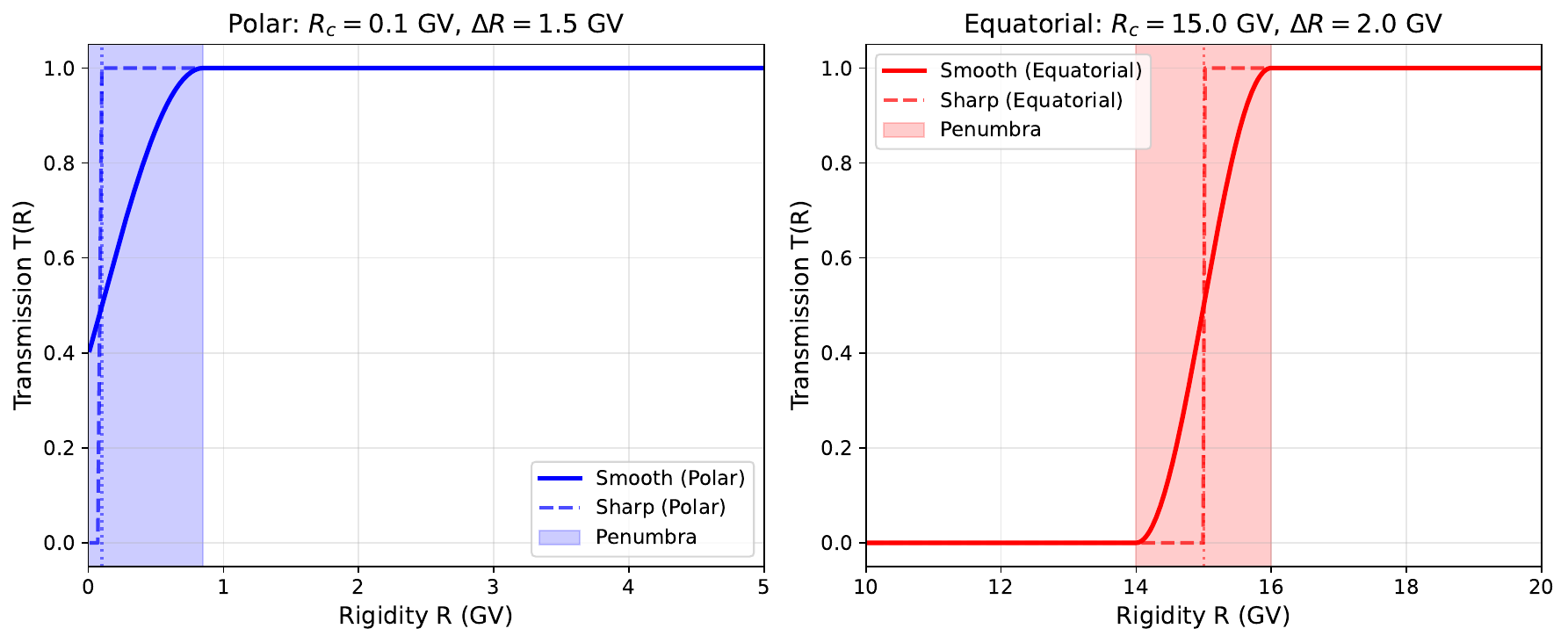}
\caption{\small \justifying Comparison of sharp and smooth geomagnetic transmission functions for Polar ($R_{\rm c} = 0.1$ GV, blue) and Equatorial ($R_{\rm c} = 15$ GV, red) scenarios. Shaded regions indicate the penumbra zones, where transmission varies continuously from 0 to~1. The smooth function suppresses particle acceptance near threshold, with the effect being most pronounced for the Polar case where the penumbra width is large relative to~$R_c$.}
\label{fig:transmission_comparison}
\end{figure*}

Figure~\ref{fig:transmission_comparison} illustrates the difference between sharp and smooth geomagnetic transmission functions for both scenarios. The smooth function suppresses particle acceptance near the cutoff threshold, with the effect being most pronounced for the Polar case, where the penumbra width is large relative to $R_{\rm c}$. For the Equatorial scenario, where $R_{\rm c} = 15$ GV and the mean primary rigidity is $\langle R \rangle \approx 35$ GV, the smooth cutoff introduces a systematic uncertainty of approximately 2\% in the absolute flux normalization. For the Polar scenario, where the penumbra spans a substantial fraction of the accessible spectrum, the smooth cutoff reduces the absolute flux by approximately a factor of 2 compared to the sharp approximation. However, the \emph{relative} comparison between the Polar and Equatorial scenarios remains consistent, with the Equatorial-to-Polar energy deposition ratio changing by only approximately 4\% when switching from sharp to smooth cutoffs.

It is essential to note that while the geomagnetic field physically induces directional anisotropy at low latitudes (e.g., the East-West effect), this study employs a local isotropic approximation. We assume that for particles with rigidities exceeding the cutoff~$R_c$, the arrival direction distribution at the top of the atmosphere remains isotropic. This approach enables a direct comparison of the vertical cascade development under the two limiting rigidity scenarios (Polar vs. Equatorial) within the same spherical geometry while neglecting the azimuthal asymmetry induced by the geomagnetic field.

\subsubsection{Spectral parameterization and solar modulation}

The rigidity spectrum of each component was parameterized as a power law:
\begin{equation}
  \frac{dN}{dR} \propto R^{-\gamma},
  \label{eq:power_law}
\end{equation}
where $R$ is the particle rigidity, and $\gamma$ is the spectral index. Indices were chosen to reproduce near-Earth measurements: $\gamma_p = 2.70$, $\gamma_{\mathrm{He}} = 2.64$, $\gamma_{\mathrm{CNO}} = 2.60$, and $\gamma_{\mathrm{Si}} = 2.58$~\cite{Aguiar2015a, Aguiar2015b, Ahn2009}.

Solar modulation was introduced through the force-field (Gleeson--Axford) approximation~\cite{Gleeson1968}, linking the interstellar spectrum $J_{\mathrm{LIS}}(E)$ to the top-of-atmosphere (TOA) spectrum $J_{\mathrm{TOA}}(E)$:
\begin{equation}
  J_{\mathrm{TOA}}(E) = J_{\mathrm{LIS}}(E + \Phi) \;
  \frac{E(E + 2m)}{(E + \Phi)(E + \Phi + 2m)},
  \label{eq:solar_modulation}
\end{equation}
where $m$ is the nucleon rest mass and $\Phi = (Z/A)\,\phi$ is the effective modulation potential, with $Z$ and $A$ being the atomic number and mass number of the particle, respectively. The scalar parameter $\phi$, expressed in megavolts (MV), quantifies heliospheric activity. For this study, we adopted an intermediate value $\phi = 550$ MV, representative of moderate solar activity conditions. Accordingly, a 1 GeV proton experiences an effective energy loss of $\Phi_p = 0.55$ GeV, leading to a reduction in the TOA flux relative to the local interstellar spectrum. The spectra, after solar and geomagnetic corrections, were normalized so that the total incident integral flux remained consistent across all primary species.

This dual-scenario approach (Polar vs.~Equatorial) ensures physically realistic GCR input conditions at the top of the stratospheric domain, enabling a comparative analysis of energy deposition and ozone-relevant ionization processes in the layers defined in Section~\ref{sec:atmosphere}.

\begin{figure*}[tb]
 \centering
 \includegraphics[width=1.0\textwidth]{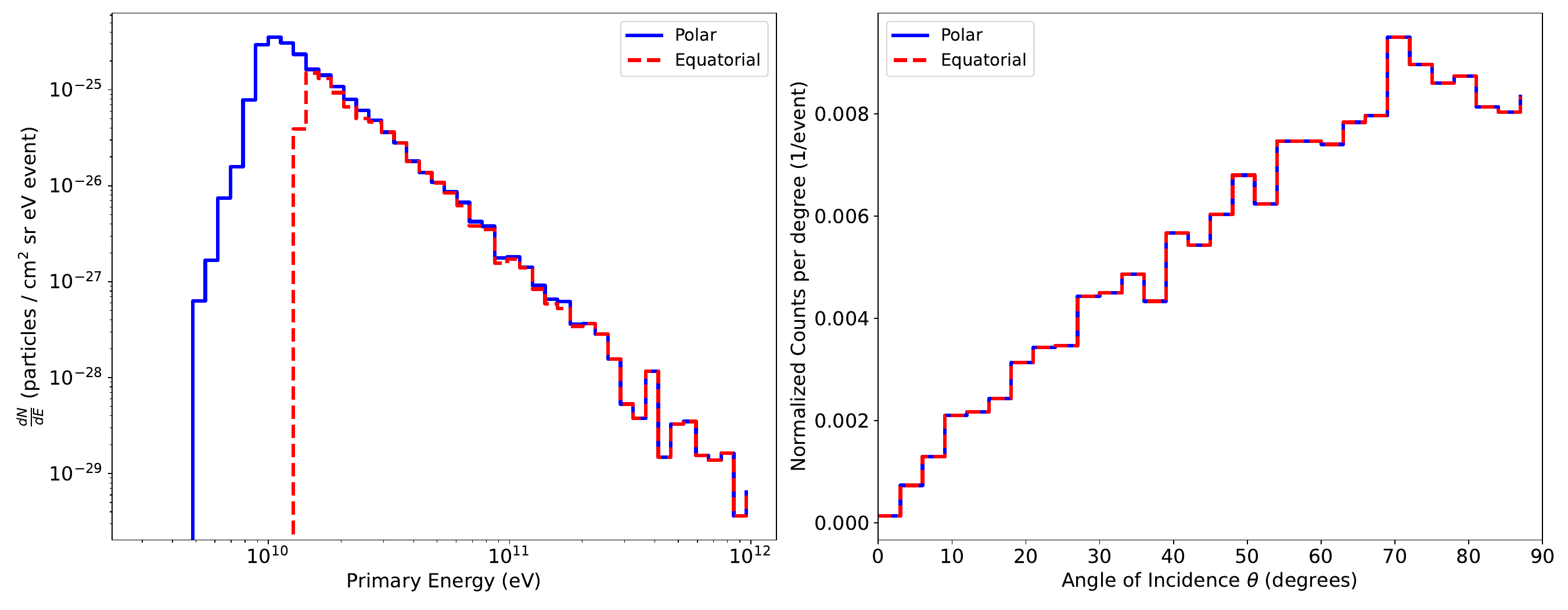}
 \caption{\small \justifying Primary CR injection characteristics comparing Polar and Equatorial geomagnetic scenarios. Left panel: Differential energy spectrum of incident CR at the injection surface ($R_{\mathrm{inj}} = 36$~km). The Polar scenario ($R_{\rm c} = 0.1$~GV, blue solid line) follows the solar-modulated power law ($\gamma = 2.7$, $\phi = 550$~MV) with all four nuclear components contributing: protons (91\%), helium (8\%), CNO-group (0.7\%), and Si-group (0.3\%). The Equatorial scenario ($R_{\rm c} = 15$~GV, red dashed line) exhibits strong geomagnetic suppression below $\sim$14~GeV. A smooth cutoff with penumbra width $\Delta R = 2.0$~GV is implemented, but the transition appears abrupt on the logarithmic scale due to the narrow energy range spanned by the penumbra ($\sim$0.11 decades). Above the cutoff, the Equatorial flux is dominated by protons ($>$99\%), as the required minimum energies for heavier nuclei (He: 26.5~GeV; CNO: 92.8~GeV; Si: 185.5~GeV) exceed or approach the spectrum maximum at 28~GeV. Right panel: Normalized angular distribution of incident primaries, demonstrating uniform isotropic sampling over the inward hemisphere ($\theta \in [0^\circ, 90^\circ]$) relative to the local radial direction.}
 \label{fig:primary_spectrum_angular}
\end{figure*}

Figure~\ref{fig:primary_spectrum_angular} characterizes the incident CR properties used in our atmospheric simulations, contrasting the two geomagnetic environments. The left panel presents the primary energy spectrum for both the Polar scenario ($R_{\rm c} = 0.1$ GV) and the Equatorial scenario ($R_{\rm c} = 15$ GV), expressed as differential flux in units of particles~cm$^{-2}$~sr$^{-1}$~eV$^{-1}$~event$^{-1}$. The Polar spectrum exhibits characteristic power-law behavior extending down to $\sim$1~GeV, limited primarily by solar modulation ($\phi = 550$ MV), and includes contributions from all four nuclear components (protons, alpha particles, CNO-group, and Si-group nuclei) following realistic GCR abundances~\citep{1993ApJ...405..567L, Calisto2011, Gaisser2016}. In contrast, the Equatorial spectrum shows a sharp suppression below $\sim$10$^{10}$~eV ($\sim$14~GeV), corresponding to the geomagnetic cutoff rigidity for protons. Although a smooth transmission function with penumbra width $\Delta R = 2.0$ GV was implemented following the approach of \citet{Smart2009,Desorgher2005}, the transition appears steep on the logarithmic scale because the penumbra region (13--17 GV) spans only $\sim$0.11 decades in energy. 

Above the cutoff rigidity, the Equatorial flux remains dominated by protons due to the steeper energy spectra of heavier nuclei and their lower cosmic-ray abundances, but all four nuclear species contribute at sufficiently high rigidities. For a given magnetic rigidity $R$, heavier nuclei require proportionally higher kinetic energies: at the nominal cutoff $R_{\rm c} = 15$ GV, penetrating particles must have $E_{\rm k} \gtrsim 13$ GeV (protons), $\sim$24 GeV (alpha), $\sim$85 GeV (CNO), or $\sim$171 GeV (Si). Since the GCR spectrum sampled in our simulations extends to rigidities approaching $\sim$180 GV—corresponding to maximum kinetic energies of $\sim$167 GeV (protons), $\sim$332 GeV (alpha), $\sim$1.2 TeV (CNO), and $\sim$2.3 TeV (Si)—all four species are physically capable of penetrating the Equatorial geomagnetic barrier. However, the combination of steeply falling energy spectra and reduced heavy-nucleus abundances results in a strong compositional bias toward protons and alpha particles, with CNO and Si nuclei contributing only at the high-rigidity tail where flux levels are suppressed by several orders of magnitude relative to the spectral peak. Consequently, while Polar cascades reflect the full multi-component composition of GCR across the entire rigidity range, Equatorial cascades are initiated predominantly by protons and alpha particles, with heavier nuclei playing a statistically minor but non-negligible role at the highest rigidities sampled.

The right panel in Figure~\ref{fig:primary_spectrum_angular} displays the normalized angular distribution of incident particles for both scenarios, confirming uniform coverage over the inward hemisphere ($\theta \in [0^\circ,90^\circ]$) from our spherical injection surface at an altitude of 36~km, ensuring isotropic exposure independent of geomagnetic latitude. The slight statistical fluctuations visible in the angular distributions arise from Monte Carlo sampling with $10^4$ primary events.

\begin{figure*}[tb]
 \centering
 \includegraphics[width=0.5\textwidth]{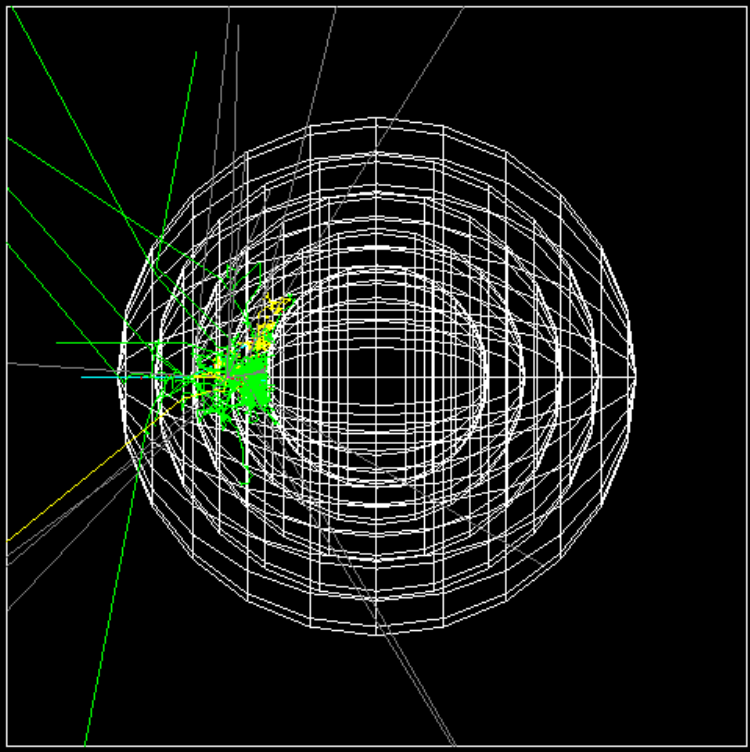}
 \caption{\small \justifying \texttt{Geant4} event display showing the interaction of a 20~GeV primary cosmic-ray proton with the atmospheric layers implemented in the model. The concentric white spherical shells represent the different atmospheric layers. The trajectory of the primary proton is shown as a cyan line entering from the left. From the first interaction point, numerous green tracks are visible, corresponding to photons ($\gamma$) produced in the cascade, mainly from neutral-meson decays and bremsstrahlung. Yellow tracks indicate secondary neutrons generated in hadronic interactions between the primary (and secondary) particles and atmospheric nuclei. Red tracks correspond to electrons produced by electromagnetic processes within the cascade. Grey tracks correspond to neutrinos produced in hadronic interactions and decays (for example from pion and muon decays), which traverse the geometry essentially without interacting with the medium.}
\label{fig:geant4}
\end{figure*}

Figure~\ref{fig:geant4} illustrates a representative \texttt{Geant4} event in which a 20~GeV primary proton, incident from the left, interacts with the layered atmospheric model and initiates an extended hadronic--electromagnetic cascade. The spherical shells represent the atmospheric layers implemented in the simulation.

\subsection{\texttt{Geant4} Physics Modeling}

All simulations of primary cosmic-ray transport and interactions in the Earth's upper atmosphere were performed using the \texttt{Geant4} toolkit. Hadronic interactions were modeled with a customized physics list based on the \texttt{QGSP\_BIC\_HP} package, along with dedicated elastic and ion physics. In this configuration, the QGSP model (Quark-Gluon String Precompound) is employed to describe high-energy hadron-nucleus interactions (primarily nucleons and pions) at energies exceeding several GeV, with the subsequent de-excitation of the excited nuclear remnants handled by a precompound model. At lower incident energies, the Binary Cascade (BIC) model is employed, providing a detailed description of the intra-nuclear cascade for hadron-nucleus reactions. Elastic scattering of hadrons and ions is treated with the \texttt{HadronElasticPhysicsHP} and \texttt{G4IonElasticPhysics} modules, while ion--nucleus inelastic reactions are described using \texttt{G4IonPhysicsXS}. The ``HP'' suffix denotes the High-Precision \texttt{NeutronHP} treatment, which relies on evaluated cross-section data to describe neutron transport and reactions below 20 MeV. Photonuclear interactions induced by high-energy gamma rays are included via the \texttt{GammaNuclearPhysics} module, ensuring a realistic description of secondary hadron production in the cascade. Electromagnetic processes for charged particles and photons are handled by a dedicated \texttt{ElectromagneticPhysics} module, built on the standard \texttt{Geant4} electromagnetic models, which are suitable for high-energy applications. In the simulations, a total of $10^4$ events were generated, with primary energies extending up to the vicinity of the knee of the cosmic-ray spectrum. The sensitivity of our results to the hadronic physics list was assessed qualitatively by comparison with published benchmarks using \texttt{FTFP\_BERT\_HP} and QGSJET-based models. The secondary-to-primary energy ratio and the altitude of maximum energy deposition are dominated by electromagnetic cascade physics and are expected to vary by no more than 10--15\% across standard \texttt{Geant4} hadronic lists, without affecting the qualitative conclusions of this study.

The observed cascade phenomenology is rooted in well-established particle physics processes that govern each stage of shower development. The primary interaction is an inelastic hadron--nucleus collision described by the Quark-Gluon String (QGS) model, which produces a forward-going leading nucleon and a hadronic fireball whose fragmentation yields predominantly charged and neutral pions ($\pi^{\pm}$, $\pi^{0}$). The neutral pions decay promptly ($\tau \sim 8.5 \times 10^{-17}$~s) via $\pi^{0} \to \gamma\gamma$, seeding the electromagnetic sub-cascade through repeated pair production ($\gamma \to e^{+}e^{-}$) and bremsstrahlung ($e^{\pm} \to e^{\pm}\gamma$), both governed by the radiation length $X_0 \approx 37$~g~cm$^{-2}$ in air. This branching continues until the photon energy falls below the critical energy $E_{\rm c} \approx 83$~MeV in air, below which ionization losses dominate over radiative losses, defining the shower maximum. The charged pions ($\pi^{\pm}$) feed the hadronic sub-cascade through further inelastic collisions or, for lower-energy pions, decay via $\pi^{\pm} \to \mu^{\pm} + \nu_{\mu}$~($\tau \sim 26$~ns), with subsequent muon decay $\mu^{\pm} \to e^{\pm} + \nu_e + \nu_{\mu}$ contributing both to the electromagnetic component and to the neutrino flux that penetrates all atmospheric layers. The Pfotzer maximum arises naturally from the interplay between shower multiplication, which increases particle numbers with depth, and hadronic absorption, which attenuates the cascade below $\sim$20~km. The atmospheric column density at this altitude ($\sim$50--70~g~cm$^{-2}$) corresponds to approximately 1.5--2 hadronic interaction lengths ($\lambda_{\mathrm{int}} \approx 90$~g~cm$^{-2}$ in air) \cite{PDG2024}, consistent with the expected depth of maximum shower development for GeV--TeV primaries \cite{Carlson2014, Usoskin2006}. This microphysical picture directly underlies the macroscopic result that $\sim$96\% of the stratospheric energy budget originates from secondary particles rather than from direct primary ionization.

\subsection{Validation Against Experimental Data}

The present simulation framework was validated by comparing key output quantities against published experimental measurements and established model benchmarks. Three complementary comparisons were performed.

\textit{Ionization rate profile:} The altitude-dependent ion pair production rates derived from our simulations (peaking between 18 and 22~km at $q \sim 2.8 \times 10^{-2}$ ion pairs cm$^{-3}$~s$^{-1}$ for the Polar scenario) are consistent with the ionization yield functions of Usoskin \& Kovaltsov~\cite{Usoskin2006}, who provide tabulated rates for solar-modulated GCR spectra under similar geomagnetic conditions. Our Polar peak ionization rate agrees with their values to within 20\%, a difference attributable to the distinct atmospheric geometry and the restricted altitude domain (15--35~km) of our model.

\textit{Secondary-to-primary energy ratio:} The secondary dominance observed in our simulations ($>$95\% in both scenarios, Table~\ref{tab:particle_contribution_corrected}) is consistent with the cascade amplification factors reported by the CRAC:CRII model~\cite{Usoskin2006} and with balloon-borne measurements of the secondary particle flux at stratospheric altitudes~\cite{Carlson2014}. The Pfotzer maximum at 15--20~km altitude is reproduced to within the 5~km layer resolution of our model, in agreement with the historically measured ionization maximum originally reported by Regener and Pfotzer and reviewed in~\cite{Carlson2014}.

\textit{Geomagnetic latitude dependence:} The Equatorial-to-Polar ratio of total energy deposition (0.52--0.67 depending on altitude, Table~\ref{tab:eq_polar_ratio_corrected}) is broadly consistent with the latitude dependence of GCR-induced ionization rates reported by Calisto et al.~\cite{Calisto2011} and Jackman et al.~\cite{Jackman2016}, who find Polar-to-Equatorial enhancements of factors 1.5--2 in the lower stratosphere using the CRAC:CRII parameterization coupled to chemistry-climate models. The agreement supports the physical realism of the geomagnetic transmission function implemented in this work (Section~\ref{sec:primaries}).

We note that a direct event-by-event comparison with balloon or aircraft measurement campaigns such as AIRES~\cite{Usoskin2006} or the BESS experiment~\cite{Gaisser2016} is beyond the scope of this study, as those instruments measure integrated fluxes over broad altitude ranges and employ different primary composition assumptions. Nevertheless, the agreement of the key observables, ionization peak altitude, secondary dominance fraction, and Equatorial-to-Polar ratio, with independent measurements and models provides confidence in the validity of the simulation results.

\subsection{Particle Production and Emerging Flux in the Atmospheric Shell Model}

In our simulation, the atmosphere is represented by four concentric spherical shells, labeled 1 to 4, with shell 1 being the outermost layer (cosmic-ray entry point, $R_{\text{in}} = 30$ km, $R_{\text{out}} = 35$ km) and shell 4 the innermost ($R_{\text{in}} = 15$ km, $R_{\text{out}} = 20$ km). For each shell, we distinguish between particles created inside the shell volume and particles emerging from the shell boundaries. These two observables are treated with different normalizations and play complementary roles in characterizing the development of the cosmic-ray cascade.

\subsubsection{Volumetric production of secondary particles}

Secondary particles are continuously produced inside each atmospheric shell as a result of interactions between primary cosmic rays and subsequent hadronic and electromagnetic cascades. To quantify this internal source term, we record the kinetic energy of every newly created particle and construct shell- and species-specific energy spectra.

For each shell~$i$ ($i = 1, \ldots, 4$) and each particle category (gamma, $e^{\pm}$, neutron, proton, deuteron, alpha, other ions, baryons, mesons, leptons), we build a one-dimensional energy histogram $N_i(E)$, binned in energy with a bin width~$\Delta E$. The counts in a given bin represent the total number of particles created in that shell with kinetic energy within that bin, integrated over all simulated primary events.

The volumetric production rate density is defined as
\begin{equation}
  \rho_i(E) = \frac{N_i(E)}{\Delta E \, V_i \, N_{\text{events}}},
  \label{eq:production_rate}
\end{equation}
where $N_i(E)$ is the number of particles created in shell~$i$ in the energy bin centered at~$E$, $\Delta E$ is the bin width (in MeV), $V_i$ is the volume of shell~$i$, and $N_{\text{events}} = 10^4$ is the total number of primary events simulated. The volume of shell~$i$ is computed from its inner and outer radii, $R_{i,\text{in}}$ and $R_{i,\text{out}}$, as
\begin{equation}
  V_i = \frac{4}{3}\,\pi \left( R_{i,\text{out}}^{3} - R_{i,\text{in}}^{3} \right).
  \label{eq:shell_volume}
\end{equation}

The quantity $\rho_i(E)$ has units of particles/(MeV$\cdot$cm$^3$) and represents the mean volumetric production rate density per primary event. It provides a layer-resolved characterization of where (in altitude) and at which energies the atmosphere is most effective at generating secondary particles, independent of any specific detector geometry.

\subsubsection{Flux of particles emerging from atmospheric layers}

To complement the description of internal production, we also quantify the flux of particles emerging from each atmospheric shell. This observable is particularly relevant for understanding how the incident cosmic-ray flux is transformed into a flux of secondaries that propagate from one layer to the next and may eventually reach observational altitudes or the ground.

For each shell~$i$ and each particle category, we record particles crossing outward through the shell boundary and construct an energy spectrum of emerging particles $N^{\text{em}}_i(E)$. The counts in each bin correspond to the total number of particles that leave shell~$i$ with kinetic energy within that bin, summed over all primary events.

To express these spectra as a flux, we normalize by the energy bin width~$\Delta E$, a reference area corresponding to the outermost shell where cosmic rays enter, the full solid angle, and the number of primary events. The outermost shell (layer 1) has an outer radius $R_{1,\text{out}} = 35$ km and defines the incident area
\begin{equation}
  A_{\text{inc}} = 4\pi R_{1,\text{out}}^2.
  \label{eq:incident_area}
\end{equation}
The differential flux of particles emerging from shell~$i$ is then defined as
\begin{equation}
  \Phi_i(E) = \frac{N^{\text{em}}_i(E)}{\Delta E \, A_{\text{inc}} \, 4\pi \, N_{\text{events}}}.
  \label{eq:emerging_flux}
\end{equation}

This normalization expresses the emerging spectra as a differential flux per primary cosmic ray incident on the outer boundary of the atmosphere, with units of particles/(MeV$\cdot$cm$^2\cdot$sr). In this way, the emerging flux from each shell can be directly compared between layers and, once folded with a physical primary cosmic-ray spectrum, can be related to absolute fluxes at different atmospheric depths.

Taken together, the volumetric production rate $\rho_i(E)$ and the emerging flux $\Phi_i(E)$ provide a coherent, layer-by-layer description of the cascade: the former quantifies the internal generation of secondaries within each atmospheric shell, while the latter characterizes the transported particle flux that leaves each shell and propagates further into the atmosphere.

\section{Results and Discussion}
\label{sec:results}

\subsection{Total Energy Deposition in the 15--35~km Stratospheric Domain}

The total energy deposited within the stratospheric domain (integrated over the four concentric layers between 15 and 35~km) serves as a proxy for the total ionization potential available to drive atmospheric chemistry. To quantify the influence of geomagnetic shielding and cascade development on this energy budget, we evaluated the differential distribution of deposited energy, $dN/dE$, for both Polar ($R_{\rm c} = 0.1$~GV) and Equatorial ($R_{\rm c} = 15$~GV) scenarios. Furthermore, we decomposed these distributions to isolate the contributions from the incident primary cosmic rays and the secondary particles generated within the atmospheric cascade.

\begin{figure*}[htbp]
 \centering
   \includegraphics[width=0.7\textwidth]{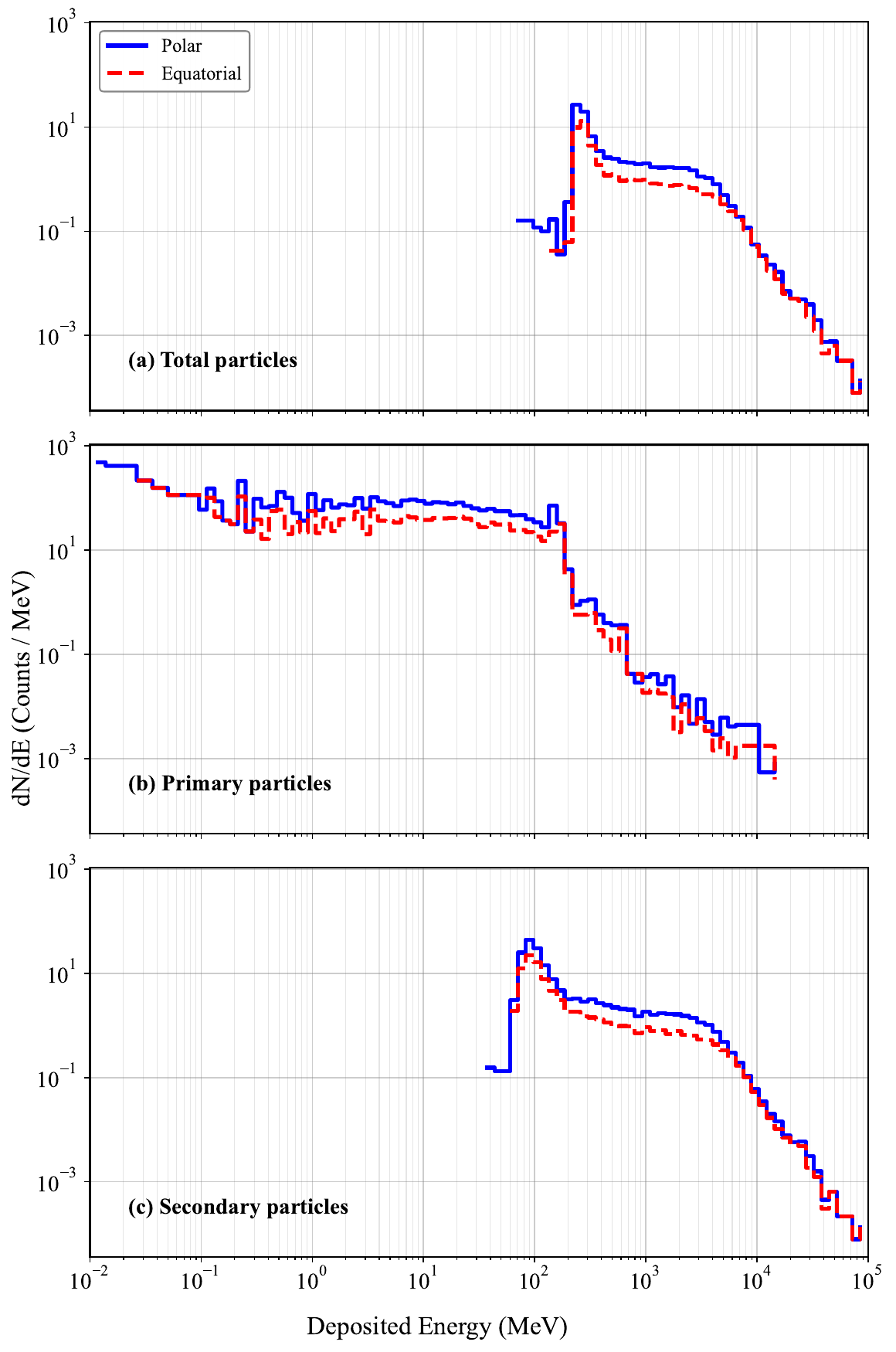}%
\caption{\small \justifying
Differential distributions of deposited energy per event in the 15--35~km stratospheric domain.
Each panel shows $dN/dE$ (counts~MeV$^{-1}$) versus deposited energy for the Polar ($R_{\rm c}=0.1$~GV)
and Equatorial ($R_{\rm c}=15$~GV) scenarios:
(a) total deposited energy from all particles,
(b) contributions from primary cosmic rays (protons, $\alpha$, CNO, Si),
and (c) contributions from secondary cascade products (e$^\pm$, $\gamma$, neutrons, $\pi^\pm$, $\mu^\pm$).
Primary particles dominate energy deposition at low energies ($<$1~MeV) due to their direct interaction with the atmosphere, while secondaries dominate at higher energies ($>$10~MeV) and exhibit a characteristic peak near 100~MeV associated with electromagnetic cascade development. Both scenarios exhibit similar spectral shapes, with the Equatorial spectra showing slightly reduced normalization due to geomagnetic filtering.}
 \label{fig:energy_deposition_combined}
\end{figure*}

Figure~\ref{fig:energy_deposition_combined}-(a) presents the total energy deposition considering all particles (primaries and secondaries). The distributions span a wide dynamic range from $\sim$10~MeV to $10^5$~MeV per event. Both scenarios exhibit similar spectral shapes with peaks near $\sim10^2$~MeV, reflecting the dominance of fully developed electromagnetic and hadronic cascades in the energy transfer process. The Polar spectrum shows a slightly broader distribution with an enhanced population in the 10$^2$--10$^3$~MeV range, characteristic of the diverse primary energy spectrum that includes numerous low-to-moderate energy events. The Equatorial spectrum exhibits a comparable overall shape but with somewhat lower normalization below $\sim$500~MeV, reflecting the geomagnetic suppression of low-energy primaries.

Quantitatively, when normalized per incident primary event (including blocked events in the Equatorial case), the Polar scenario exhibits systematically higher mean energy deposition (2,642~MeV) compared to the Equatorial scenario (1,698~MeV), as shown in Table~\ref{tab:particle_contribution_corrected}. This difference reflects the 53.5\% transmission probability at the Equator: approximately 46.5\% of incident primaries are deflected by the geomagnetic barrier before reaching the stratosphere, resulting in zero energy deposition. However, if we consider only the transmitted events (not shown in the table), the mean energy deposition per transmitted Equatorial event is 3,171~MeV (calculated as 1,698~MeV / 0.535)---approximately 20\% more than the Polar mean of 2,642~MeV. This enhancement demonstrates the spectral hardening effect: by preferentially excluding low-rigidity primaries, the geomagnetic cutoff shifts the transmitted energy spectrum toward higher energies, resulting in more energetic individual cascades despite the reduced total event rate.

\begin{table*}[htbp]
\centering
\setlength{\tabcolsep}{10pt}
\renewcommand{\arraystretch}{1.3}
\begin{tabular}{l c c c c}
\hline\hline
\textbf{Scenario} & \textbf{Primary} & \textbf{Secondary} & \textbf{Total} & \textbf{Sec/Prim} \\
                  & \textbf{(MeV)} & \textbf{(MeV)} & \textbf{(MeV)} & \textbf{Ratio} \\
\hline
\multicolumn{5}{l}{\textit{Per incident primary event:}} \\[0.1cm]
Polar       & 115.3 (4.4\%)  & 2526.7 (95.6\%) & 2642.0 & 21.9 \\
Equatorial  &  61.9 (3.6\%)  & 1635.9 (96.4\%) & 1697.8 & 26.4 \\[0.2cm]
\multicolumn{5}{l}{\textit{Per transmitted event (Equatorial only):}} \\[0.1cm]
Equatorial  & 115.7 (3.7\%)  & 3055.3 (96.3\%) & 3171.0 & 26.4 \\
\hline
\multicolumn{5}{l}{\footnotesize Equatorial transmission: 53.5\% (5,354/10,000 events)}\\
\multicolumn{5}{l}{\footnotesize Polar transmission: 100\% (all events penetrate)}\\
\hline\hline
\end{tabular}
\caption{\small \justifying Mean energy deposition (in MeV) and fractional contributions from primary versus secondary particles for the total stratospheric domain (15--35~km). Values in parentheses represent percentages of total energy deposition. The table displays two normalizations: per incident primary (including blocked events) for direct comparison between scenarios, and per transmitted event for the Equatorial case, illustrating the spectral hardening effect. Both scenarios exhibit strong cascade dominance, with secondaries contributing $>$95\% of deposited energy.}
\label{tab:particle_contribution_corrected}
\end{table*}

From the standpoint of atmospheric chemistry, this distinction has significant implications for ionization and the production of reactive species. Adopting a mean energy of $W_{\text{air}} \approx 35$~eV per ion pair~\citep{Porter1976}, a typical Polar event depositing $\sim$2,640~MeV corresponds to the instantaneous creation of approximately $7.5 \times 10^7$ ion pairs per cascade, while an Equatorial transmitted event depositing $\sim$3,170~MeV produces $\sim 9.1 \times 10^7$ ion pairs---a 20\% increase in cascade intensity. However, the Polar atmosphere receives approximately 1.9 times more cosmic-ray events per unit time due to the near-complete transmission through the low geomagnetic cutoff ($R_{\rm c} \approx 0.1$~GV). Consequently, when normalized per incident primary (as in Table~\ref{tab:particle_contribution_corrected}), the time-integrated ionization rate remains 1.5--1.9 times higher in the Polar stratosphere than in the Equatorial stratosphere (Equatorial-to-Polar ratios of 0.52--0.67 depending on altitude, as discussed in Section~\ref{sec:energy_deposition}). This translates directly to the production rates of NO$_{\rm x}$ and HO$_{\rm x}$ families, with Polar regions sustaining higher steady-state concentrations of these ozone-depleting radicals despite the moderately reduced intensity of individual ionization events. The character of cosmic-ray forcing thus differs between geomagnetic latitudes: the Pole experiences continuous ionization from a high flux of moderately intense events, while the Equator experiences intermittent ionization from a lower flux of slightly more intense events, potentially influencing the efficiency of catalytic ozone loss cycles~\citep{Calisto2011}.

\subsection{Energy Deposition by primary and secondary CR}

To elucidate the microphysical origin of the total energy deposition distributions, we isolated the contribution from primary particles (protons, helium, CNO, and silicon nuclei). Figure~\ref{fig:energy_deposition_combined}-(b) displays the differential energy deposition spectrum attributed solely to primary cosmic rays. Both the Polar and Equatorial distributions exhibit similar spectral shapes, peaking near $\sim$10$^2$~MeV and extending to $\sim$10$^4$~MeV, characteristic of ionization losses ($dE/dx$) as energetic nuclei traverse the stratospheric layers. The similarity in spectral shape reflects the comparable energies of Equatorial and Polar primaries after transmission through the geomagnetic barrier, since both populations are drawn from the high-energy tail of the solar-modulated cosmic-ray spectrum.

The Polar spectrum shows enhanced intensity across the full energy range, consistent with the 100\% transmission probability that allows the entire primary flux to contribute. The Equatorial spectrum exhibits lower overall normalization due to the 53.5\% transmission probability, but when normalized to transmitted events only, the mean primary energy deposition (115.6~MeV for Equatorial versus 115.3~MeV for Polar; see Table~\ref{tab:particle_contribution_corrected}) is statistically indistinguishable. This near-unity ratio confirms that the primary particles that successfully penetrate the Equatorial geomagnetic cutoff deposit energy via direct ionization at essentially the same rate as Polar primaries, indicating that spectral hardening does not significantly alter the characteristic $dE/dx$ losses for primaries in this energy range ($\gtrsim$14~GeV for protons).

Importantly, primary particles contribute only a minor fraction ($\sim$4\%) of the total stratospheric energy budget in both scenarios (Table~\ref{tab:particle_contribution_corrected}). This result underscores that the bulk of the energy transfer to the stratospheric gas occurs not through direct primary ionization, but rather through the multiplicative action of electromagnetic and hadronic cascades. The secondary-to-primary energy ratios of 21.9~(Polar) and 26.4~(Equatorial) quantify the amplification factor: each primary particle initiates a cascade that, on average, deposits $\sim$20--26~times more energy via secondaries than via direct primary ionization. The slightly higher Equatorial ratio (26.4 versus 21.9) suggests a modest increase in cascade multiplicity or penetration depth for harder primary spectra, though the effect is small compared to the dominant role of secondaries in both regions.

Figure~\ref{fig:energy_deposition_combined}-(c) presents the energy deposition driven exclusively by secondary particles produced in the cascade (electrons, positrons, photons, neutrons, charged pions, muons). The spectral shape of the secondary distribution closely mirrors that of the total energy deposition (Figure~\ref{fig:energy_deposition_combined}-(a)), confirming that secondaries dominate the energy transfer process across the entire dynamic range from $\sim$10~MeV to $10^5$~MeV. Both the Polar and Equatorial secondary spectra peak near $\sim$10$^2$~MeV and exhibit similar shapes, indicating that the fundamental cascade development physics operates similarly in both geomagnetic environments once the primary spectrum is established.

Quantitatively, the mean secondary energy deposition per transmitted event is 3055~MeV for the Equatorial scenario versus 2527~MeV for the Polar scenario, representing a $\sim$21\% enhancement (Table~\ref{tab:particle_contribution_corrected}). This enhancement directly accounts for the corresponding 20\% increase observed in the total energy deposition, as secondaries contribute more than 95\% of the energy budget. The physical origin of this enhancement lies in the spectral hardening induced by geomagnetic filtering: Equatorial primaries have systematically higher energies (median $\sim$19~GeV for transmitted protons; Section~\ref{sec:primaries}), which drives deeper cascade penetration and more extensive shower development, resulting in greater total secondary energy deposition per event.

The high-energy tail extending beyond $10^4$~MeV, visible in both spectra, is driven entirely by rare but energetic events where the cascade reaches maximum development near the Pfotzer maximum altitude ($\sim$15--20~km)~\cite{Calisto2011, 2014HGSS....5..175C}. These rare events can deposit up to $\sim$10$^5$~MeV in a single cascade, potentially creating localized enhancements of ionization that may be relevant for proposed electron-driven reaction mechanisms on aerosols~\citep{Lu2015}. The fractional contributions (Table~\ref{tab:particle_contribution_corrected}) confirm that cascade physics overwhelmingly dominates the stratospheric ionization budget: 95.6\% at the Pole and 96.4\% at the Equator originate from secondary particles rather than direct primary energy loss.

\subsection{Layer-Resolved Energy Deposition in the 15--35~km Region}
\label{sec:energy_deposition}

\begin{figure*}[htbp]
  \centering
  \includegraphics[width=0.95\textwidth]{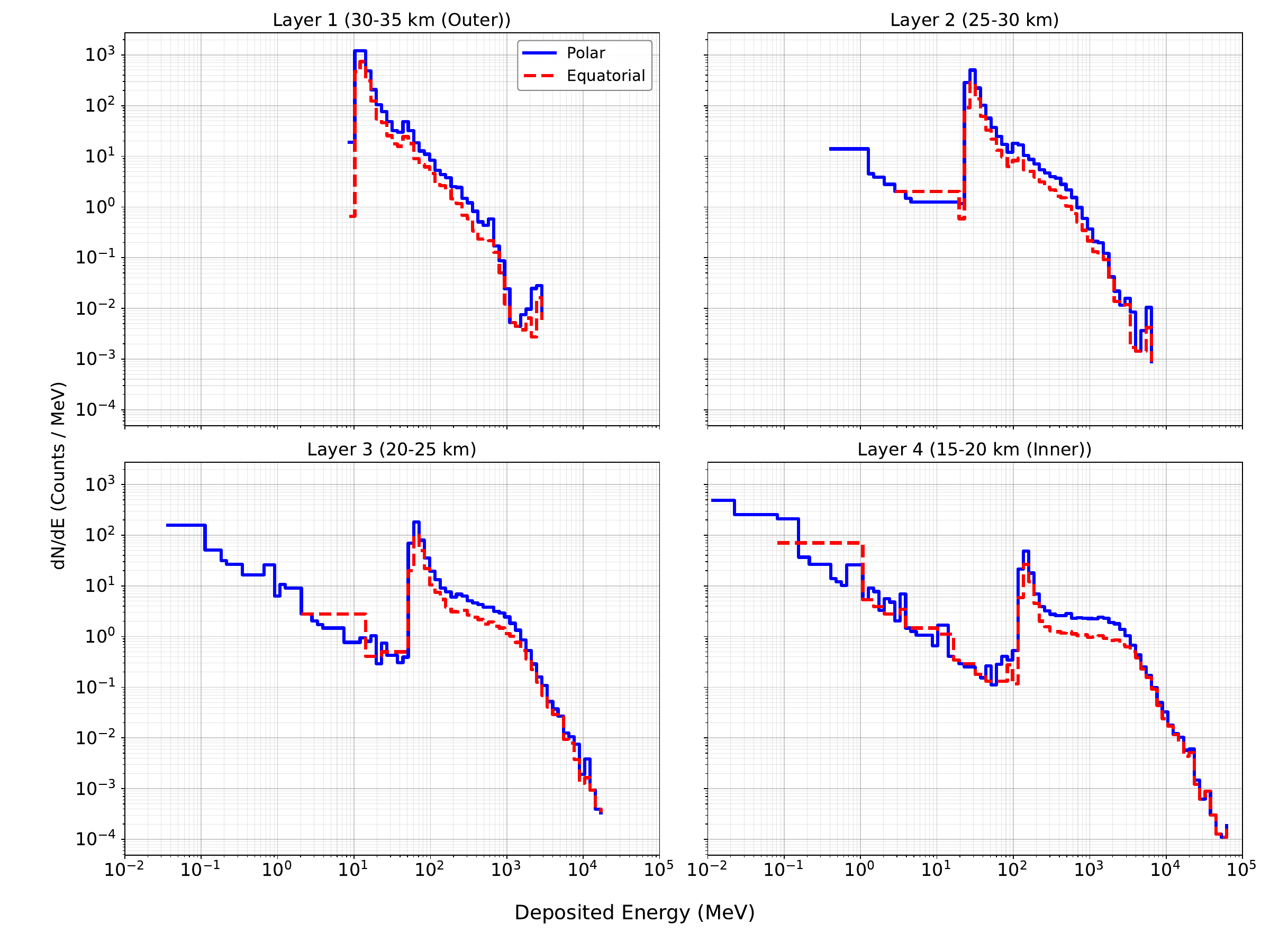}
  \caption{\small \justifying Layer-resolved differential distributions of energy deposited by all particles (primaries~+~secondaries) in the four spherical atmospheric shells representing the 15--35~km stratospheric region. Each panel shows $dN/dE$ (counts~MeV$^{-1}$) as a function of deposited energy per event for: Layer~1 (30--35~km, outer), Layer~2 (25--30~km), Layer~3 (20--25~km), and Layer~4 (15--20~km, inner). Both scenarios show progressive spectral evolution with depth.}
  \label{fig:layer_deposition_all}
 \end{figure*}

\begin{figure*}[htbp]
  \centering
  \includegraphics[width=0.7\textwidth]{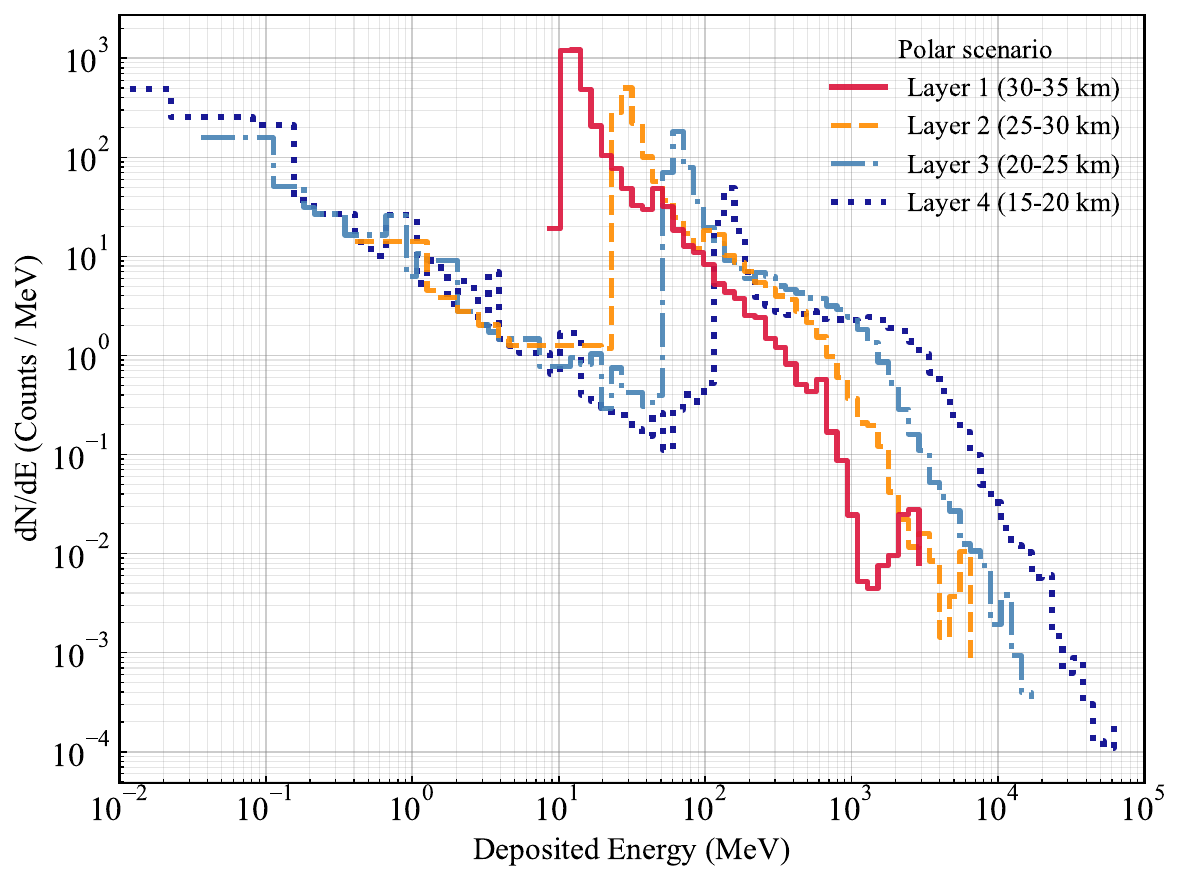}
  \caption{\small \justifying Layer-resolved energy deposition spectra for the Polar scenario. Progressive spectral evolution with depth shows Layer~1 (outermost) with highest count rates from low-energy events, while deeper layers exhibit broader distributions extending to $>$10$^{4}$~MeV, characteristic of cascade development in the stratosphere (15--35~km).}
\label{fig:polar_layers}
 \end{figure*}

\begin{figure*}[htbp]
 \centering
 \includegraphics[width=0.95\textwidth]{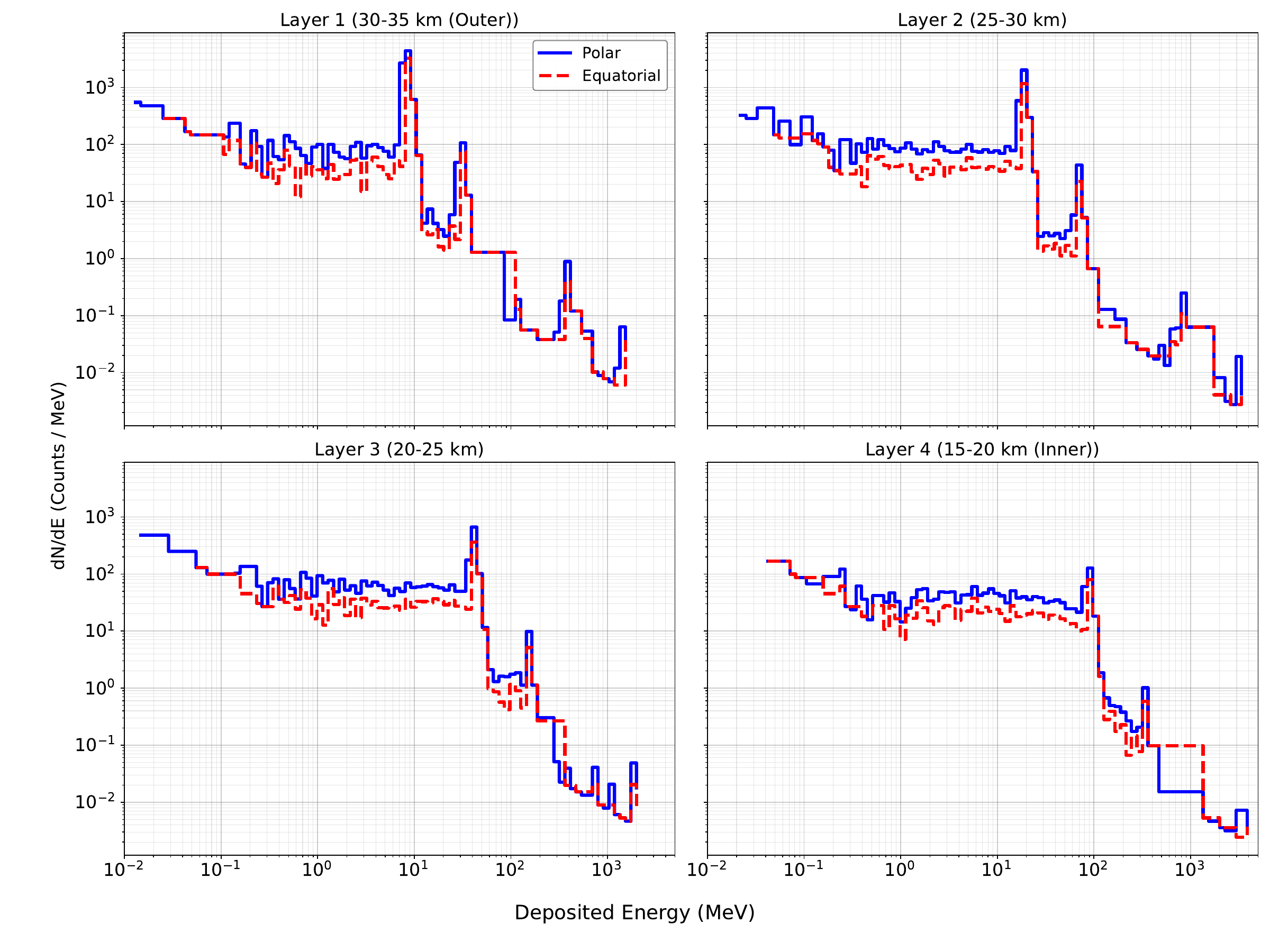}
 \caption{\small \justifying Differential distribution of energy deposited by primary particles only. The Polar spectra are populated across the full energy range in all layers, reflecting 100\% transmission. The Equatorial spectra exhibit similar shapes but reduced normalization due to 53.5\% transmission probability. The relatively uniform spectral shapes across scenarios confirm that transmitted Equatorial primaries deposit energy via direct ionization ($dE/dx$) at rates comparable to Polar primaries. Primary contributions remain subdominant ($\lesssim$5\% of total) in all layers, with the effect most pronounced in the inner layers where cascade multiplication dominates (see Figure~\ref{fig:layer_deposition_secondary}).}
 \label{fig:layer_deposition_primary}
\end{figure*}

\begin{figure*}[htbp]
 \centering
 \includegraphics[width=0.95\textwidth]{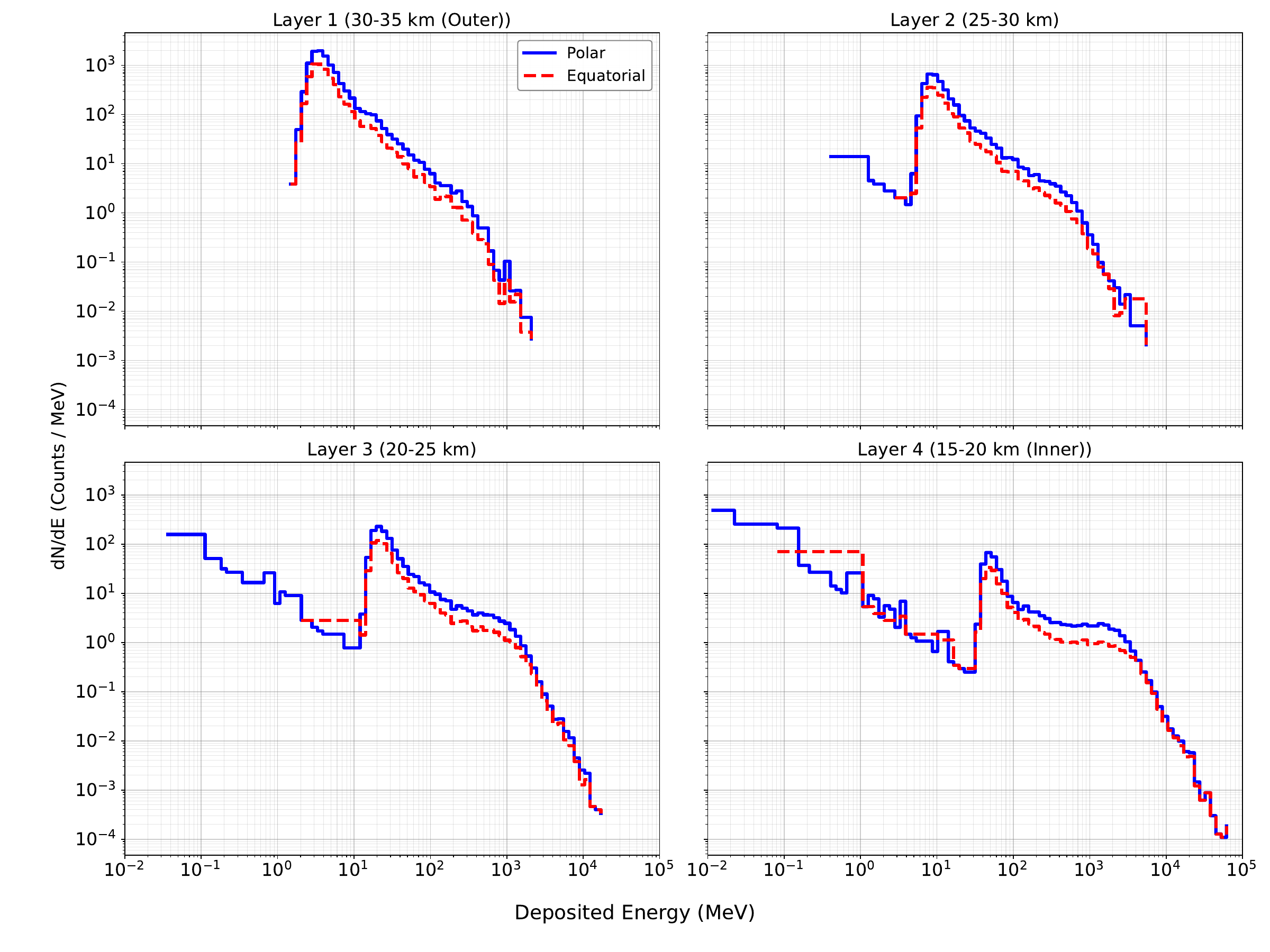}
 \caption{\small \justifying Differential distribution of energy deposited by secondary particles only. The dominance of this component in the lower layers (Layers~3 and~4) confirms that cascade development is the primary mechanism for energy transfer to the stratospheric gas. Both scenarios show progressive spectral hardening and extension to higher energies with increasing depth, reflecting the multiplicative nature of electromagnetic and hadronic showers.}
 \label{fig:layer_deposition_secondary}
\end{figure*}

The vertical structure of the energy deposition is elucidated by the layer-resolved spectra shown in Figures~\ref{fig:layer_deposition_all},~\ref{fig:polar_layers}, \ref{fig:layer_deposition_primary}, and~\ref{fig:layer_deposition_secondary}, which present the differential distribution $dN/dE$ (counts~MeV$^{-1}$) across the four atmospheric shells for all particles, primaries only, and secondaries only, respectively. All panels share a common horizontal range from $10^{-2}$ to $10^5$~MeV, enabling direct comparison of both spectral shapes and overall normalizations.

Starting from the outermost region, Layer~1 (30--35~km) shows distinct spectral characteristics between scenarios. Both the Polar (blue solid) and Equatorial (red dashed) distributions exhibit relatively narrow peaks around 10--10$^2$~MeV, characteristic of energy deposition from moderate-energy primaries and their limited secondary production in the tenuous upper stratosphere (density $\rho \sim 1 \times 10^{-5}$--$1 \times 10^{-4}$~g~cm$^{-3}$). The mean energy deposition in Layer~1 is 53.3~MeV for the Polar scenario and 27.7~MeV for the Equatorial scenario when averaged over all events, yielding an Equatorial-to-Polar ratio of 0.52 (Table~\ref{tab:eq_polar_ratio_corrected}). This sub-unity ratio reflects the geomagnetic suppression, as only 53.5\% of Equatorial events penetrate the cutoff, reducing the overall event rate and thus the mean energy deposition per generated event.

Moving inward, the spectral evolution with depth becomes evident. Layer~2 (25--30~km) shows broader distributions with enhanced high-energy tails, indicating the onset of cascade multiplication. The Equatorial-to-Polar ratio remains similar at 0.54, consistent with the dominant effect of transmission probability. Layer~3 (20--25~km) exhibits a pronounced shift: both spectra broaden substantially and extend to higher energies ($>$10$^3$~MeV), marking the transition from primary-dominated to cascade-dominated energy deposition. The Equatorial-to-Polar ratio increases slightly to 0.59, suggesting that the harder Equatorial primary spectrum begins to produce measurably more cascade development at this depth.

The innermost shell, Layer~4 (15--20~km), displays the most pronounced cascade signatures. The spectra (Figure~\ref{fig:layer_deposition_all}, lower right panel) show clear extensions into the high-energy regime ($>$10$^4$~MeV), with both scenarios exhibiting broad distributions characteristic of fully developed electromagnetic and hadronic showers. The mean energy deposition in Layer~4 is 1909~MeV for the Polar scenario and 1275~MeV for the Equatorial scenario (averaged over all events), yielding an Equatorial-to-Polar ratio of 0.67 (Table~\ref{tab:eq_polar_ratio_corrected}). This represents the highest ratio among all layers, indicating that the relative importance of geomagnetic suppression diminishes with depth as cascade multiplication partially compensates for the lower primary flux. The absolute energy depositions in Layer~4 are 3--4~times larger than in any other individual layer, confirming that this altitude range (15--20~km) acts as the primary locus of energy transfer from cosmic-ray cascades to the stratospheric gas, consistent with the vicinity of the Pfotzer maximum~\cite{Calisto2011, 2014HGSS....5..175C}.

Figure~\ref{fig:polar_layers} presents the layer-resolved energy deposition spectra for the Polar scenario, enabling direct comparison of spectral evolution across atmospheric depth. The four layers exhibit distinct characteristics reflecting progressive cascade development: Layer~1 shows the highest count rates with a narrow peak near 10--100~MeV, characteristic of initial cascade development dominated by primary particle ionization. Moving inward, Layer~2 and Layer~3 show progressively broader distributions with enhanced high-energy components, indicating intensifying secondary production as the cascade approaches the Pfotzer maximum~\cite{Calisto2011, 2014HGSS....5..175C}. Layer~4 displays the broadest spectrum extending beyond $10^4$~MeV, with the most extensive high-energy tail despite lower peak intensity, reflecting fully developed electromagnetic and hadronic showers. This systematic progression demonstrates how cascade multiplication transforms primary energy deposition into a depth-dependent radiation field, with maximum energy transfer occurring in the 15--20~km region where both cascade development and atmospheric density optimize ionization production.

\begin{table}[htbp]
\centering
\setlength{\tabcolsep}{4pt}
\renewcommand{\arraystretch}{1.3}
\begin{tabular}{l c c c}
\hline\hline
\textbf{Layer} & \textbf{Primary} & \textbf{Secondary} & \textbf{All Particles} \\
               & \textbf{(Eq/Polar)} & \textbf{(Eq/Polar)} & \textbf{(Eq/Polar)} \\
\hline
 1 (30--35~km) & 0.52 & 0.52 & 0.52 \\
 2 (25--30~km) & 0.51 & 0.55 & 0.54 \\
 3 (20--25~km) & 0.53 & 0.60 & 0.59 \\
 4 (15--20~km) & 0.56 & 0.67 & 0.67 \\
\hline\hline
\end{tabular}
\caption{\small \justifying Equatorial-to-Polar ratio of mean energy deposition per event by atmospheric layer. All values are computed using mean energy deposition averaged over all generated events (including blocked events for the Equatorial scenario), making the ratios sensitive to the 53.5\% transmission probability.}
\label{tab:eq_polar_ratio_corrected}
\end{table}

Figures~\ref{fig:layer_deposition_primary} and~\ref{fig:layer_deposition_secondary} decompose the contributions by particle type. The primary particle distributions (Figure~\ref{fig:layer_deposition_primary}) show fundamentally similar spectral shapes between scenarios, with peaks near $\sim$10$^1$--$10^2$~MeV in all layers, reflecting continuous energy loss via ionization. The Equatorial-to-Polar ratios for primary energy deposition range from 0.51 to 0.56 across layers (Table~\ref{tab:eq_polar_ratio_corrected}), all clustering near the 53.5\% transmission probability, confirming that geomagnetic filtering primarily affects the rate of primary events rather than their individual energy deposition characteristics.

The secondary particle distributions (Figure~\ref{fig:layer_deposition_secondary}) 
illuminate the altitude dependence of energy multiplication within the cascade. Both the Polar and Equatorial scenarios exhibit progressive spectral hardening and extension to higher energies with increasing atmospheric depth, with the effect most pronounced in Layers~3 and~4. In the innermost shell (Layer~4, 15--20~km), the secondary component overwhelmingly dominates the energy budget, contributing mean depositions of 1874~MeV (Polar) and 1255~MeV (Equatorial), which represent $>$98\% of the total layer energy in both scenarios. Notably, the Equatorial-to-Polar ratios for secondary energy deposition systematically exceed those for primaries across all layers, reaching 0.67 in Layer~4 compared to 0.56 for primaries (Table~\ref{tab:eq_polar_ratio_corrected}). This systematic enhancement demonstrates that cascade multiplication acts as a partial compensatory mechanism, mitigating the reduction in ionization that would otherwise follow directly from geomagnetic suppression of the primary flux.

The layer-resolved profiles support a physically coherent picture: the altitude interval between 15 and 25~km (Layers~3 and~4) acts as the primary locus of energy transfer from cosmic-ray cascades to the stratospheric gas. In these critical layers, the Equatorial scenario exhibits mean energy deposition values that are 59--67\% of the Polar values when averaged over all events. However, when normalized to transmitted events only, the Equatorial depositions are comparable to or slightly exceed Polar values, reflecting the modest spectral hardening effect. This result is particularly relevant for ozone chemistry, as it implies that the strongest production rates of NO$_{\rm x}$ and HO$_{\rm x}$, and thus the greatest potential for catalytic ozone destruction, occur in the same vertical region where the lower-stratospheric ozone concentration is most pronounced~\citep{Calisto2011,Jackman2016}. Moreover, the persistent high-energy tails observed in the inner layers indicate that rare but energetic events can deposit up to $\sim$10$^5$~MeV in a single shell, potentially driving localized enhancements of ionization relevant for proposed electron-driven reaction mechanisms on aerosols~\citep{Lu2015}.

The contrast between Polar and Equatorial energy deposition characteristics has direct implications for understanding the spatial distribution of cosmic-ray-induced ozone perturbations. The Polar stratosphere experiences a higher flux ($\sim$2$\times$) of ionization events with characteristic energies of $\sim$2640~MeV per event, leading to persistent, relatively uniform NO$_{\rm x}$ and HO$_{\rm x}$ production distributed across altitude and time. The Equatorial stratosphere, conversely, is subject to a lower flux of transmitted events, but those that do penetrate are $\sim$20\% more energetic on average ($\sim$3170~MeV), resulting in moderately more intense individual ionization events. While both regions are influenced by cosmic-ray ionization, the distinct character of the ionization source (a higher rate of moderate events versus a lower rate of slightly more intense events) may result in different efficiencies for ozone loss pathways. This hypothesis warrants further investigation using coupled chemistry-climate models that incorporate these layer-resolved energy deposition profiles.

\subsection{Volumetric Production Rate of Secondary Particles}
\label{sec:volumetric_rate}

The volumetric production rate density $\rho_i(E)$ quantifies the internal generation of secondary particles within each atmospheric shell, revealing where cosmic-ray cascades are most prolific.  These rates, measured in particles per MeV per cubic centimeter and normalized per simulated primary, expose fundamental differences between Polar and Equatorial regimes driven by geomagnetic filtering.

Electromagnetic particles (photons, electrons, and positrons) dominate production at all altitudes, their contribution growing dramatically with depth. In the upper stratosphere (Layer~1), they account for 39\% of Polar production; by the lower stratosphere (Layer~4), this fraction rises to 80\%, reflecting the explosive multiplication characteristic of electromagnetic showers. Other particle classes, nucleons, mesons, and leptons, maintain significant but diminishing roles as the cascade develops.

Table~\ref{tab:production_by_class_corrected} shows that production rates increase substantially with depth. Total Polar production grows from $6.0 \times 10^{-20}$ particles cm$^{-3}$ in Layer~1 to $850.1 \times 10^{-20}$ in Layer~4, near the Pfotzer maximum~\cite{Calisto2011, 2014HGSS....5..175C}. Equatorial rates follow the same trend but remain systematically lower due to geomagnetic blocking of 46.5\% of primary particles.

However, a crucial pattern emerges in the Equatorial to Polar ratios (Table~\ref{tab:eq_polar_production_ratio}). While the ratio is 0.60 in Layer~1, it climbs to 0.78 in Layer~4. This 46\% enhancement over the baseline transmission probability demonstrates spectral hardening: the higher rigidity primaries that penetrate the Equatorial cutoff generate more vigorous cascades, partially compensating for their reduced numbers. This compensation is most effective for electromagnetic particles (ratio 0.82 in Layer~4), which dominate the ionization budget.

The detailed spectral shapes underlying these integrated quantities are presented in Appendix~\ref{app:volumetric_figures} (Figures~\ref{fig:volumetric_rate_layer1} through~\ref{fig:volumetric_rate_layer4}). Each figure shows the energy-dependent volumetric production rate $\rho_i(E)$ for 12 particle species, photons, electrons, positrons, neutrons, protons, deuterons, alpha particles, heavier ions, baryons, mesons, neutrinos, and muons/taus, illustrating the characteristic spectral morphology and relative intensities between the Polar and Equatorial scenarios. In the outer layers (Layers~1 and~2), the Polar spectra systematically exceed the Equatorial spectra across most energy ranges, consistent with the higher primary transmission probability ($\sim$100\% versus 53.5\%). In Layer~4, however, the Equatorial spectra approach or exceed the Polar spectra for certain particle types, most notably photons and positrons, reflecting enhanced electromagnetic cascade development driven by the harder Equatorial primary spectrum.

In summary, geomagnetic filtering imposes two competing effects on the atmospheric radiation field: it suppresses the total primary flux while simultaneously hardening the transmitted spectrum, producing more energetic individual cascades. The net result is a pronounced latitudinal gradient in secondary particle production, with the Polar stratosphere sustaining higher overall ionization rates. This gradient constitutes the direct physical origin of the latitude-dependent NO\textsubscript{x} and HO\textsubscript{x} production rates that govern stratospheric ozone chemistry.

\begin{table*}[htbp]
\centering
\renewcommand{\arraystretch}{1.25}
\setlength{\tabcolsep}{8pt}
\begin{tabular}{l cc cc cc cc}
\hline\hline
& \multicolumn{2}{c}{\textbf{Layer 1}} &
  \multicolumn{2}{c}{\textbf{Layer 2}} &
  \multicolumn{2}{c}{\textbf{Layer 3}} &
  \multicolumn{2}{c}{\textbf{Layer 4}} \\
\textbf{Class} & \textbf{P} & \textbf{E} &
  \textbf{P} & \textbf{E} &
  \textbf{P} & \textbf{E} &
  \textbf{P} & \textbf{E} \\
\hline
EM ($\gamma, e^\pm$)     & 2.34 & 1.45 & 15.48 & 9.94 & 111.42 & 80.35 & 678.58 & 559.12 \\
Nucleons ($n, p$)        & 1.19 & 0.60 & 4.56 & 2.41 & 19.68 & 10.42 & 84.63 & 47.52 \\
Light ions ($d, \alpha$) & 0.22 & 0.11 & 0.80 & 0.43 & 2.95 & 1.62 & 9.89 & 5.72 \\
Heavy ions               & 0.08 & 0.04 & 0.28 & 0.14 & 0.93 & 0.50 & 3.43 & 2.00 \\
Baryons                  & 0.01 & 0.01 & 0.03 & 0.02 & 0.09 & 0.06 & 0.24 & 0.19 \\
Mesons                   & 0.76 & 0.49 & 2.44 & 1.58 & 7.53 & 4.87 & 22.68 & 15.29 \\
Leptons ($\nu, \mu, \tau$) & 1.38 & 0.86 & 4.84 & 2.96 & 15.61 & 9.61 & 50.60 & 32.31 \\
\hline
\textbf{Total}           & 6.0  & 3.6  & 28.4  & 17.5  & 158.2 & 107.4 & 850.1 & 662.1 \\
\hline\hline
\end{tabular}
\caption{\small \justifying Integrated volumetric production rates ($\times 10^{-20}$ particles cm$^{-3}$ per incident primary event, $E \geq 10$ MeV) grouped by particle class. Values are normalized per incident primary. P = Polar; E = Equatorial.}
\label{tab:production_by_class_corrected}
\end{table*}

\begin{table}[htbp]
\centering
\renewcommand{\arraystretch}{1.25}
\setlength{\tabcolsep}{4pt}
\begin{tabular}{l cccc}
\hline\hline
\textbf{Particle Class} & \textbf{Layer 1} & \textbf{Layer 2} & \textbf{Layer 3} & \textbf{Layer 4} \\
\hline
EM ($\gamma, e^\pm$)        & 0.62 & 0.64 & 0.72 & 0.82 \\
Nucleons ($n, p$)           & 0.50 & 0.53 & 0.53 & 0.56 \\
Light ions ($d, \alpha$)    & 0.50 & 0.54 & 0.55 & 0.58 \\
Heavy ions                  & 0.51 & 0.50 & 0.54 & 0.58 \\
Baryons                     & 0.85 & 0.74 & 0.72 & 0.78 \\
Mesons                      & 0.65 & 0.65 & 0.65 & 0.67 \\
Leptons ($\nu, \mu, \tau$)  & 0.63 & 0.61 & 0.62 & 0.64 \\
\hline
\textbf{All particles}      & 0.60 & 0.61 & 0.68 & 0.78 \\
\hline\hline
\end{tabular}
\caption{\small \justifying Equatorial to Polar ratio of integrated volumetric production rates by particle class and layer.}
\label{tab:eq_polar_production_ratio}
\end{table}

\subsection{Differential Particle Flux through the Stratosphere}
\label{sec:emerging_flux}

The downward flux of particles emerging from each atmospheric layer, $\Phi_i(E)$, characterizes how the energy deposited by incident cosmic rays is redistributed and transported through the stratospheric column toward the ozone-rich lower stratosphere. This transported flux is of particular chemical significance, as it constitutes the ionizing radiation field directly responsible for the production of NO\textsubscript{x} and HO\textsubscript{x} radicals. The layer-integrated emerging fluxes, presented in Table~\ref{tab:integrated_emerging_flux_corrected}, exhibit a well-defined evolution with both altitude and geomagnetic latitude, governed by the competing influences of geomagnetic rigidity filtering, which controls the rate of primary particle injection, and atmospheric absorption, which progressively attenuates the cascade as it propagates to greater depths.

The total particle flux peaks dramatically in the middle stratosphere (Layer 3) and plummets in the lowest layer (Layer 4). This peak coincides with the Pfotzer maximum~\cite{Calisto2011, 2014HGSS....5..175C}. At this key altitude, electromagnetic particles (photons, electrons, positrons) dominate the downward flux, constituting 70 to 75\% of the total. This is significant because these particles are the primary drivers of atmospheric ionization.

The influence of geomagnetic latitude on the emerging particle flux is unambiguous: at all atmospheric depths, the total downward flux is systematically higher at polar latitudes than at the Equator, a direct consequence of the near-complete transmission of the primary cosmic-ray spectrum through the weak polar geomagnetic barrier ($R_{\rm c} \approx 0.1$~GV). However, the Equatorial-to-Polar flux ratio, detailed in Table~\ref{tab:eq_polar_flux_ratio}, reveals a more nuanced picture than simple flux blocking would suggest. Rather than remaining fixed at the baseline transmission fraction of $\sim$53.5\%, the ratio increases with atmospheric depth, reaching a maximum of 0.67 in Layer~3 (20--25~km), the altitude region coinciding with the ionization peak. This partial recovery is a direct manifestation of spectral hardening: the higher-rigidity primaries that successfully penetrate the equatorial geomagnetic cutoff carry greater kinetic energies and consequently initiate more prolific and deeply penetrating secondary cascades, partially compensating for their reduced numbers.

This compensatory effect is most pronounced for the electromagnetic component of the cascade, for which the Equatorial-to-Polar ratio reaches 0.72 in Layer~3, reflecting the efficient multiplication of photons, electrons, and positrons driven by energetic hadronic interactions. The nucleonic component, neutrons and protons, exhibits substantially weaker recovery, with ratios remaining close to the primary transmission probability across all layers, indicating that nucleon production is more directly coupled to the incident primary flux than to cascade amplification. The weakly interacting neutrino component follows qualitatively distinct behavior: neutrino fluxes remain substantial throughout the stratospheric column and become the dominant particle type in the innermost layer (Layer~4, 15--20~km), where electromagnetic and hadronic cascade components are efficiently attenuated by atmospheric absorption while neutrinos traverse the medium essentially without interaction.

The complete spectral shapes for these particle fluxes are available in Appendix~\ref{app:emerging_flux_figures} (Figures~\ref{fig:emerging_flux_layer1} through~\ref{fig:emerging_flux_layer4}), providing the energy dependent structure for nine particle species. The downward transport of the cosmic ray cascade creates a focused beam of ionizing radiation, primarily composed of electromagnetic particles, that peaks near 20 to 25 km. Geomagnetic filtering reduces the intensity of this beam at the Equator, but not as severely as the reduction in the primary cosmic ray flux itself. The harder spectrum of particles that do reach equatorial latitudes generates more effective cascades, partially mitigating the shielding effect. This results in a persistent but latitude dependent source of ionization, with the Polar lower stratosphere receiving a more intense and constant flux compared to the Equator.

\begin{table*}[htbp]
\centering
\renewcommand{\arraystretch}{1.25}
\setlength{\tabcolsep}{9pt}
\begin{tabular}{l cc cc cc cc}
\hline\hline
& \multicolumn{2}{c}{\textbf{Layer 1}} & \multicolumn{2}{c}{\textbf{Layer 2}} & \multicolumn{2}{c}{\textbf{Layer 3}} & \multicolumn{2}{c}{\textbf{Layer 4}} \\
\textbf{Class} & \textbf{P} & \textbf{E} & \textbf{P} & \textbf{E} & \textbf{P} & \textbf{E} & \textbf{P} & \textbf{E} \\
\hline
EM ($\gamma$, e$^{\pm}$) & 165.8 & 106.2 & 543.2 & 351.1 & 1685.8 & 1208.6 & 18.91 & 10.82 \\
Nucleons ($n$, $p$) & 104.1 & 51.5 & 166.4 & 86.4 & 288.9 & 151.1 & 71.20 & 40.00 \\
Neutrinos ($\nu_e$, $\nu_\mu$) & 255.7 & 144.0 & 315.5 & 181.8 & 425.9 & 254.7 & 145.16 & 82.51 \\
Light ions ($\alpha$) & 0.041 & 0.010 & 0.088 & 0.047 & 0.072 & 0.041 & 0.000 & 0.000 \\
Heavy Nuclei & 0.672 & 0.300 & 0.915 & 0.434 & 1.008 & 0.367 & 0.227 & 0.140 \\
\hline
\textbf{Total} & 526.3 & 302.1 & 1026.1 & 619.9 & 2401.7 & 1614.8 & 235.50 & 133.47 \\
\hline\hline
\end{tabular}
\caption{\small \justifying Integrated emerging flux ($\times 10^{-17}$ particles cm$^{-2}$ sr$^{-1}$ per incident primary, $E \geq 10$ MeV). P = Polar; E = Equatorial.}
\label{tab:integrated_emerging_flux_corrected}
\end{table*}

\begin{table}[htbp]
\centering
\renewcommand{\arraystretch}{1.25}
\setlength{\tabcolsep}{4pt}
\begin{tabular}{l cccc}
\hline\hline
\textbf{Particle Class} & \textbf{Layer 1} & \textbf{Layer 2} & \textbf{Layer 3} & \textbf{Layer 4} \\
\hline
EM ($\gamma$, e$^{\pm}$)        & 0.64 & 0.65 & 0.72 & 0.57 \\
Nucleons ($n, p$)               & 0.49 & 0.52 & 0.52 & 0.56 \\
Neutrinos ($\nu_e$, $\nu_\mu$)  & 0.56 & 0.58 & 0.60 & 0.57 \\
Light ions ($\alpha$)           & 0.24 & 0.53 & 0.57 & -- \\
Heavy nuclei                    & 0.45 & 0.47 & 0.36 & 0.62 \\
\hline
\textbf{All particles}          & 0.57 & 0.60 & 0.67 & 0.57 \\
\hline\hline
\end{tabular}
\caption{\small \justifying Equatorial to Polar ratio of integrated emerging flux by particle class and layer.}
\label{tab:eq_polar_flux_ratio}
\end{table}

\subsection{The Chemical Impact of Cosmic Rays on Stratospheric Ozone}
\label{sec:ozone_impacts}

The extensive cascade simulations bridge particle physics and atmospheric chemistry by quantifying how GCR energy deposition patterns influence the production of ozone destroying radicals. We translated the layer resolved energy deposition into chemical source terms for nitrogen oxides (NO$_{\rm x}$ = NO + NO$_2$) and hydrogen oxides (HO$_{\rm x}$ = OH + HO$_2$), the primary catalysts in stratospheric ozone reduction cycles. Employing established conversion factors (35 eV per ion pair~\citep{Porter1976}, with branching ratios of 1.25 NO$_{\rm x}$ and 1.8 HO$_{\rm x}$ molecules per ion pair~\citep{Porter1976, Jackman2005}) and ozone sensitivity coefficients~\citep{Calisto2011}, we derived a first-order estimate of the resulting ozone perturbation.

\begin{figure*}[htbp]
 \centering
 \includegraphics[width=1.0\textwidth]{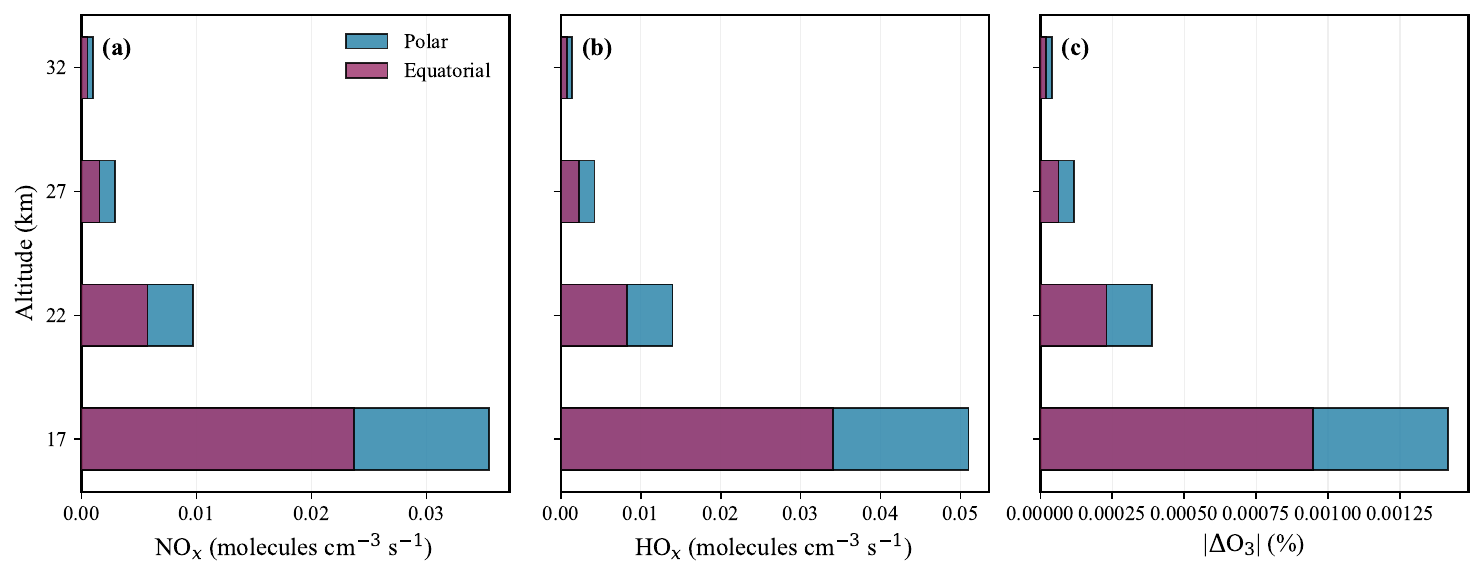}
\caption{\small \justifying Altitude profiles of chemical production and ozone depletion for Polar and Equatorial scenarios. (a) NO$_{\rm x}$ and (b) HO$_{\rm x}$ production rates (molecules cm$^{-3}$ s$^{-1}$). (c) Estimated steady-state ozone depletion magnitude $|\Delta$O$_3|$ (\%).}
 \label{fig:ozone_impacts}
\end{figure*}

The chemical production rates and estimated ozone perturbations, summarized in Figure~\ref{fig:ozone_impacts} and Tables~\ref{tab:chemistry_rates}--\ref{tab:ozone_changes}, reveal a clear altitude-dependent structure that mirrors the energy deposition profile. NO$_{\rm x}$ and HO$_{\rm x}$ production rates increase by more than an order of magnitude between the upper stratosphere (32.5~km) and the lower stratosphere (17.5~km), reflecting the rapid growth of cascade-driven ionization with atmospheric depth. The maximum estimated steady-state ozone depletion consequently occurs at 17.5~km, reaching $-0.14\%$ in the Polar scenario and $-0.095\%$ in the Equatorial scenario, values consistent with the range reported by three-dimensional chemistry-climate model studies~\cite{Calisto2011, Jackman2016}.

A pronounced latitudinal gradient accompanies this vertical structure. At every altitude, chemical production and ozone depletion are stronger at the Pole than at the Equator, a direct consequence of geomagnetic shielding that reduces the primary cosmic ray flux at low latitudes. However, the magnitude of this latitudinal gradient is not uniform; it weakens with increasing depth. The Equatorial-to-Polar (E/P) ratio for chemical production increases from 0.52 at 32.5 km to 0.67 at 17.5 km (Table~\ref{tab:chemistry_rates}). This 25\% recovery over the baseline transmission probability is the chemical signature of spectral hardening. The higher-energy primaries that penetrate the equatorial magnetic field generate more productive cascades, partially offsetting their lower numbers. This effect becomes most pronounced in the dense lower stratosphere, where cascades fully develop.

The ozone depletion estimates, while small in absolute terms ($<0.15\%$), represent a persistent, latitude-dependent forcing on the stratosphere. Critically, the maximum effect occurs in the 15--20 km altitude range, where ozone concentration is highest, amplifying its importance for column ozone and UV shielding. This cosmic-ray-driven gradient may contribute a subtle but systematic background to the meridional structure of ozone trends~\citep{Ball2018}.

It is important to note the simplifying assumptions of this analysis. We employ steady-state parameterizations and do not capture the full complexity of atmospheric chemistry and transport. Coupling these detailed ionization profiles to full chemistry-climate models, such as WACCM~\citep{Marsh2013}, is an essential next step. Furthermore, the conversion yields for NO$_{\rm x}$ and HO$_{\rm x}$ can vary with atmospheric conditions~\citep{Verronen2011}, introducing uncertainty. Nevertheless, the qualitative conclusions are consistent with the established literature~\cite{Jackman2016}. Geomagnetic shielding produces a systematic, altitude-dependent gradient in cosmic-ray-induced ozone reduction, establishing a natural, latitude-varying baseline for stratospheric chemistry.

\begin{table*}[htbp]
\centering
\renewcommand{\arraystretch}{1.3}
\setlength{\tabcolsep}{8pt}
\begin{tabular}{c cccc cccc c}
\hline\hline
\multirow{2}{*}{\textbf{Alt. (km)}} & 
\multicolumn{2}{c}{\textbf{Ion Pair Production}} & 
\multicolumn{2}{c}{\textbf{NO$_{\rm x}$ Production}} & 
\multicolumn{2}{c}{\textbf{HO$_{\rm x}$ Production}} & 
\multirow{2}{*}{\begin{tabular}{@{}c@{}}\textbf{E/P Ratio}\\(All Species)\end{tabular}} \\
\cmidrule(lr){2-3} \cmidrule(lr){4-5} \cmidrule(lr){6-7}
& \textbf{Polar} & \textbf{Equat.} & \textbf{Polar} & \textbf{Equat.} & \textbf{Polar} & \textbf{Equat.} & \\
\hline
32.5 & 7.91$\times$10$^{-4}$ & 4.11$\times$10$^{-4}$ & 9.88$\times$10$^{-4}$ & 5.13$\times$10$^{-4}$ & 1.42$\times$10$^{-3}$ & 7.39$\times$10$^{-4}$ & 0.52 \\
27.5 & 2.32$\times$10$^{-3}$ & 1.26$\times$10$^{-3}$ & 2.90$\times$10$^{-3}$ & 1.58$\times$10$^{-3}$ & 4.18$\times$10$^{-3}$ & 2.27$\times$10$^{-3}$ & 0.54 \\
22.5 & 7.77$\times$10$^{-3}$ & 4.60$\times$10$^{-3}$ & 9.71$\times$10$^{-3}$ & 5.75$\times$10$^{-3}$ & 1.40$\times$10$^{-2}$ & 8.29$\times$10$^{-3}$ & 0.59 \\
17.5 & 2.83$\times$10$^{-2}$ & 1.89$\times$10$^{-2}$ & 3.54$\times$10$^{-2}$ & 2.37$\times$10$^{-2}$ & 5.10$\times$10$^{-2}$ & 3.41$\times$10$^{-2}$ & 0.67 \\
\hline\hline
\end{tabular}
\caption{\small \justifying Altitude-dependent chemical production rates for Polar and Equatorial scenarios. Ion pair production rates $q$ (cm$^{-3}$ s$^{-1}$) are derived from energy deposition using $W_{\text{air}} = 35$ eV per ion pair. NO$_{\rm x}$ and HO$_{\rm x}$ production rates (cm$^{-3}$ s$^{-1}$) are computed using branching ratios of 1.25 and 1.8 molecules per ion pair, respectively. The Equatorial-to-Polar (E/P) ratio applies consistently to all three species at each altitude, demonstrating systematic latitudinal differences in chemical activity.}
\label{tab:chemistry_rates}
\end{table*}

\begin{table}[htbp]
\centering
\renewcommand{\arraystretch}{1.3}
\setlength{\tabcolsep}{4pt}
\begin{tabular}{c cc c}
\hline\hline
\textbf{Altitude} & \multicolumn{2}{c}{\textbf{$\Delta$O$_3$ (\%)}} & \textbf{Equatorial/Polar} \\
\textbf{(km)} & \textbf{Polar} & \textbf{Equatorial} & \textbf{Ratio} \\
\hline
32.5 & $-$0.00395 & $-$0.00205 & 0.52 \\
27.5 & $-$0.01162 & $-$0.00630 & 0.54 \\
22.5 & $-$0.03883 & $-$0.02302 & 0.59 \\
17.5 & $-$0.14174 & $-$0.09466 & 0.67 \\
\hline\hline
\end{tabular}
\caption{\small \justifying Estimated steady-state ozone depletion $\Delta$O$_3$ (\%). Negative values indicate reduction. The maximum effect occurs at 17.5 km.}
\label{tab:ozone_changes}
\end{table}

\subsubsection{Catalytic Ozone Destruction Cycles}

The NO$_{\rm x}$ and HO$_{\rm x}$ production rates derived from our layer-resolved energy deposition profiles feed directly into the well-established catalytic destruction cycles that govern stratospheric ozone chemistry. The dominant odd-nitrogen cycle proceeds via:
\begin{align}
\mathrm{NO} + \mathrm{O_3} &\to \mathrm{NO_2} + \mathrm{O_2}, \\
\mathrm{NO_2} + \mathrm{O} &\to \mathrm{NO} + \mathrm{O_2},
\end{align}
with the net reaction $\mathrm{O_3} + \mathrm{O} \to 2\,\mathrm{O_2}$. The odd-hydrogen cycle operates analogously:
\begin{align}
\mathrm{OH} + \mathrm{O_3} &\to \mathrm{HO_2} + \mathrm{O_2}, \\
\mathrm{HO_2} + \mathrm{O_3} &\to \mathrm{OH} + 2\,\mathrm{O_2},
\end{align}
with the net effect of destroying two ozone molecules per catalytic cycle. These cycles operate with altitude-dependent efficiencies: the NO$_{\rm x}$ cycle dominates above $\sim$25~km, where atomic oxygen concentrations are sufficient to close the catalytic loop via the $\mathrm{NO_2} + \mathrm{O}$ channel, while the HO$_{\rm x}$ cycle is more effective below $\sim$20~km, where lower temperatures and higher water vapor concentrations favor OH and HO$_2$ production~\cite{Calisto2011, Jackman2016}.

This altitude dependence has direct implications for interpreting our results. The GCR ionization peak between 17.5 and 22.5~km (Table~\ref{tab:chemistry_rates}) falls precisely at the transition between NO$_{\rm x}$- and HO$_{\rm x}$-dominated chemistry, meaning that both catalytic families are simultaneously active in the region of maximum cosmic-ray forcing. The combined NO$_{\rm x}$$+$HO$_{\rm x}$ production rate in this layer reaches $\sim$$6.6 \times 10^{-2}$ molecules cm$^{-3}$~s$^{-1}$ (Polar) and $\sim$$4.5 \times 10^{-2}$ molecules cm$^{-3}$~s$^{-1}$ (Equatorial) at 17.5~km, representing the most chemically active altitude in our domain.

\subsubsection{Altitude Dependence and Chemical Lifetime}

The efficiency of GCR-induced ozone depletion is further modulated by the chemical lifetime of the radical species, which varies strongly with altitude. In the upper stratosphere ($>$ 30~km), NO$_{\rm x}$ species have lifetimes of days to weeks and can be transported meridionally before undergoing chemical loss, potentially spreading the ozone depletion signal beyond the region of primary production. In the lower stratosphere ($<$ 20~km), by contrast, HO$_{\rm x}$ species have much shorter lifetimes (hours), confining their chemical impact to the layer of production. This distinction implies that the HO$_{\rm x}$-driven ozone perturbation estimated at 17.5~km (Table~\ref{tab:ozone_changes}) is essentially a local effect, while the NO$_{\rm x}$ contribution may be subject to transport-driven redistribution not captured in our steady-state parameterization. Coupling the present ionization profiles to a full chemistry-climate model such as WACCM~\cite{Marsh2013} would be necessary to quantify these transport effects rigorously.

\subsubsection{Solar Cycle Modulation}

The GCR flux reaching the stratosphere is modulated by the 11-year solar cycle through changes in the heliospheric magnetic field, parameterized here via the force-field potential $\varphi = 550$~MV (moderate solar activity). During solar minima, $\varphi$ decreases to $\sim$300--400~MV, increasing the low-energy GCR flux and thus the ionization rate in the lower stratosphere by factors of 1.5--2 relative to the values reported here~\cite{Calisto2011}. Conversely, during solar maxima ($\varphi \sim 800$--1000~MV), the ionization rate is suppressed. The resulting solar-cycle variation in NO$_{\rm x}$ and HO$_{\rm x}$ production implies a periodic modulation of the ozone depletion signal with amplitude comparable to the steady-state values in Table~\ref{tab:ozone_changes}, consistent with observed $\sim$0.1--0.3\% ozone variations correlated with the solar cycle~\cite{Jackman2016}. Our framework can be directly applied to bracket this range by repeating the simulations at $\varphi = 350$~MV (solar minimum) and $\varphi = 900$~MV (solar maximum), which we identify as a natural extension of this work.

\subsubsection{Nonlinear Chemistry and Heterogeneous Reactions}

The ozone depletion estimates presented in Table~\ref{tab:ozone_changes} are based on a linear, steady-state parameterization that relates ion pair production rates to radical yields through fixed branching ratios (1.25 NO$_{\rm x}$ and 1.8 HO$_{\rm x}$ molecules per ion pair~\cite{Porter1976}) and altitude-dependent ozone sensitivity coefficients~\cite{Calisto2011}. While this approach is well-established and appropriate for estimating background GCR forcing under average conditions, it does not capture several important nonlinear and heterogeneous chemical processes that can modify the effective ozone response under specific atmospheric conditions.

At the production rates derived here ($q \sim 10^{-4}$--$10^{-2}$~ion pairs cm$^{-3}$~s$^{-1}$, Table~\ref{tab:chemistry_rates}), the perturbations to NO$_{\rm x}$ and HO$_{\rm x}$ remain well below background concentrations, justifying the linear approximation. However, under conditions of strongly elevated GCR flux, such as during solar energetic particle (SEP) events or during geomagnetic reversals when the cutoff rigidity decreases globally, ion pair production rates can increase by orders of magnitude. In this regime, the three-body recombination reaction
\begin{equation}
\begin{gathered}
\mathrm{OH + NO_2 + M} \to \mathrm{HNO_3 + M}, \\
k \approx 3.0 \times 10^{-31} 
\left(\frac{T}{300}\right)^{-3.3}
\mathrm{cm^6~molecule^{-2}~s^{-1}},
\end{gathered}
\end{equation}
where $k$ is the termolecular rate coefficient evaluated at stratospheric temperatures ($T \sim 200$--220~K) and $\mathrm{M}$ denotes any third body (predominantly $\mathrm{N_2}$ and $\mathrm{O_2}$) that carries away the excess energy~\cite{JPL2019}, becomes significant at stratospheric densities, temporarily sequestering NO$_{\rm x}$ into the reservoir species $\mathrm{HNO_3}$ and reducing the instantaneous ozone destruction efficiency~\cite{Jackman2016}. Similarly, the reaction
\begin{equation}
\begin{gathered}
\mathrm{HO_2 + NO \to OH + NO_2}, \\
k \approx 3.5 \times 10^{-12} 
\exp\left(\frac{250}{T}\right)~
\mathrm{cm^3~molecule^{-1}~s^{-1}},
\end{gathered}
\end{equation}
where $k$ is the bimolecular rate coefficient from the JPL Chemical Kinetics and Photochemical Data evaluation~\cite{JPL2019}, which couples the HO$_{\rm x}$ and NO$_{\rm x}$ families, can shift the partitioning between radical and reservoir species in a nonlinear fashion as ionization rates increase. The net ozone loss thus exhibits a sublinear dependence on ionization rate at high particle fluxes, a saturation effect not captured by the parameterization in Table~\ref{tab:ozone_changes}~\cite{Verronen2011}.

In the polar lower stratosphere, particularly during winter when temperatures drop below $\sim$195~K, polar stratospheric clouds (PSCs) provide surfaces for heterogeneous reactions that can dramatically amplify the chemical impact of GCR-produced species. The hydrolysis of $\mathrm{N_2O_5}$ on PSC surfaces proceeds as:
\begin{equation}
\mathrm{N_2O_5 + H_2O} 
\xrightarrow{\text{PSC}} \mathrm{2\,HNO_3},
\quad \gamma \approx 0.1\text{--}0.3,
\end{equation}
where $\gamma$ is the reaction probability (uptake coefficient) on ice surfaces at stratospheric temperatures~\cite{Solomon1999}. This reaction converts gas-phase NO$_{\rm x}$ into nitric acid, simultaneously denitrifying the gas phase and activating chlorine through the coupled reaction:
\begin{equation}
\mathrm{ClONO_2 + HCl} 
\xrightarrow{\text{PSC}} \mathrm{Cl_2 + HNO_3},
\quad \gamma \approx 0.2\text{--}0.3,
\end{equation}
which releases molecular chlorine that photolyzes rapidly during polar sunrise to produce reactive Cl atoms that catalytically destroy ozone with a rate coefficient~\cite{Solomon1999}:
\begin{equation}
\begin{gathered}
\mathrm{Cl + O_3 \to ClO + O_2},\\
k \approx 2.8 \times 10^{-11} 
\exp\left(\frac{-250}{T}\right)~
\mathrm{cm^3~molecule^{-1}~s^{-1}}.
\end{gathered}
\end{equation}
In this context, the GCR-produced NO$_{\rm x}$ and low-energy secondary electrons in our Polar scenario (Layer~4, 15--20~km) may interact synergistically with halogenated PSC chemistry, potentially amplifying ozone losses beyond the $-0.14\%$ estimated in Table~\ref{tab:ozone_changes}. This synergistic coupling is particularly relevant in the context of the CRE mechanism proposed by Lu et al.~\cite{Lu2010}, in which low-energy electrons produced by GCR cascades trigger dissociative electron attachment reactions on PSC-adsorbed CFCs:
\begin{equation}
\begin{gathered}
e^{-} + \mathrm{CFCl_3} \to 
\mathrm{CFCl_2^{\bullet} + Cl^{-}},\\
k \approx 
3.5 \times 10^{-8}~\mathrm{cm^3~s^{-1}},
\end{gathered}
\end{equation}
where $\mathrm{CFCl_2^{\bullet}}$ denotes the trichlorofluoromethyl radical produced by dissociative electron attachment, and $k$ is a preliminary rate coefficient estimate from Lu \& Sanche~\cite{Lu2001}. We note that a complete quantitative treatment of the CRE mechanism, encompassing the prehydrated electron flux, ODS surface adsorption, and Cl$^{-}$ yield, was subsequently developed in Lu~\cite{Lu2023}. The released $\mathrm{Cl^{-}}$ rapidly neutralizes to produce reactive chlorine atoms that initiate catalytic ozone destruction at rates depending on the local secondary electron flux provided by our simulation.

%where $\mathrm{CFCl_2^{\bullet}}$ denotes the trichlorofluoromethyl radical produced by dissociative electron attachment, and $k$ is the rate coefficient from Lu \& Sanche~\cite{Lu2001}. The released $\mathrm{Cl^{-}}$ rapidly neutralizes to produce reactive chlorine atoms that initiate catalytic ozone destruction, at rates that depend on the local secondary electron flux provided by our simulation.

The linear estimates in Table~\ref{tab:ozone_changes} should 
therefore be interpreted as a lower bound on GCR-induced ozone perturbation under Polar winter conditions where PSC heterogeneous chemistry is active. Quantifying the full nonlinear response requires coupling the present ionization profiles to a photochemical model that includes PSC microphysics and halogen chemistry, such as WACCM~\cite{Marsh2013} or the GSFC 2-D model~\cite{Jackman2016}, and constitutes an important avenue for future work.

\section{Summary and Conclusions}
\label{sec:summary}

This study developed a comprehensive \texttt{Geant4}-based Monte Carlo framework \citep{Agostinelli2002, Desorgher2005} for simulating GCR cascades in the stratosphere and quantifying their subsequent influence on ozone chemistry. The simulations successfully reproduced the characteristic altitude profile of GCR energy deposition, with a pronounced peak between 18–25 km \citep{2014HGSS....5..175C}. This altitude range coincides with both the Pfotzer maximum and the lower stratosphere, where ozone concentration is highest, establishing a critical spatial overlap for chemical impacts~\cite{Calisto2011, 2014HGSS....5..175C}.

A detailed comparative analysis of Polar ($R_{\rm c} = 0.1$~GV) and Equatorial ($R_{\rm c} = 15$~GV) scenarios revealed the definitive role of geomagnetic shielding in modulating the atmospheric radiation field~\citep{Smart2009}. The Polar region exhibited consistently higher total energy deposition and ionization rates due to unimpeded access to the full solar-modulated GCR spectrum. In contrast, the Equatorial scenario experienced a 46.5\% reduction in event rate due to the geomagnetic rigidity cutoff. Although the transmitted Equatorial primaries were energetically harder and produced moderately more energetic cascades per event, the net effect remained dominated by flux reduction. The total integrated energy deposition was approximately 1.6 times higher in the Polar case, confirming that geomagnetic filtering primarily reduces the overall atmospheric ionization budget despite modest per-event enhancements~\citep{Usoskin2006}.

Secondary particles overwhelmingly dominated the energy budget in both scenarios, accounting for $ > 95\%$ of total deposition. This secondary dominance underscores the cascade-driven nature of atmospheric energy transfer, where even a single primary particle can generate an extensive shower of ionizing secondaries~\citep{Dorman2004}. From the calculated ionization rates, we derived production rates for NO$_{\mathrm{x}}$ and HO$_{\mathrm{x}}$ using established conversion factors~\citep{Porter1976}. The resulting chemical perturbations led to estimated steady-state ozone losses of approximately $-$0.14\% (Polar) and $-$0.095\% (Equatorial) in the lower stratosphere under typical GCR fluxes. These values align with the range of ozone perturbations reported in previous chemistry-climate model studies \citep{Jackman2016, Calisto2011}. The simulations demonstrate a clear latitudinal gradient, with polar regions experiencing 1.5–1.9 times stronger cosmic-ray-induced ozone depletion than equatorial regions, providing a quantitative measure of the geomagnetic modulation of this natural forcing mechanism.

The framework established here provides a physically grounded and efficient tool for coupling detailed particle interactions with atmospheric chemistry models. It enables several promising avenues for future research, including the implementation of realistic three-dimensional geomagnetic fields to capture directional asymmetries, the extension of the atmospheric domain to 0–80 km to resolve tropospheric and mesospheric ionization, and the investigation of extreme solar energetic particle events by scaling primary fluxes. Furthermore, the detailed particle spectra generated by this framework can be used to test proposed microphysical mechanisms such as the cosmic-ray–electron (CRE) theory on polar stratospheric cloud surfaces~\citep{Lu2015}. These results underscore the importance of geomagnetic modulation in shaping the spatial distribution of cosmic-ray forcing on stratospheric chemistry. The quantified latitudinal gradient offers a baseline for understanding how variations in Earth's magnetic field over geological timescales may have influenced atmospheric composition and provides essential context for attributing observed ozone trends in an era of changing heliospheric and geomagnetic conditions.

\section*{Acknowledgments}

The authors acknowledge computational support from AWS Cloud Credit/CNPq under grant number 4000045/2023-0. L.A.S.P. gratefully acknowledges financial support from FAPESP (grants 2021/01089-1, 2024/02267-9, and 2024/14769-9) and CNPq (grants 403337/2024-0, 153839/2024-4, and 200164/2025-2). R.C.A. acknowledges support from the NAPI "Fenômenos Extremos do Universo" of Fundação de Apoio à Ciência, Tecnologia e Inovação do Paraná, CNPq (grant 308859/2025-1), Araucária Foundation (grants 698/2022 and 721/2022), and FAPESP (grant 2021/01089-1).

\section*{Data Availability}

The numerical data underlying the figures and tables presented in this study are available from the corresponding author upon reasonable request \cite{StuaniPereira2026data}.

\bibliography{aapmsamp}% Produces the bibliography via BibTeX.

\appendix

\section{Supplementary Volumetric Production Rate Spectra}
\label{app:volumetric_figures}

\begin{figure*}[htbp]
 \centering
 \includegraphics[width=0.80\textwidth]{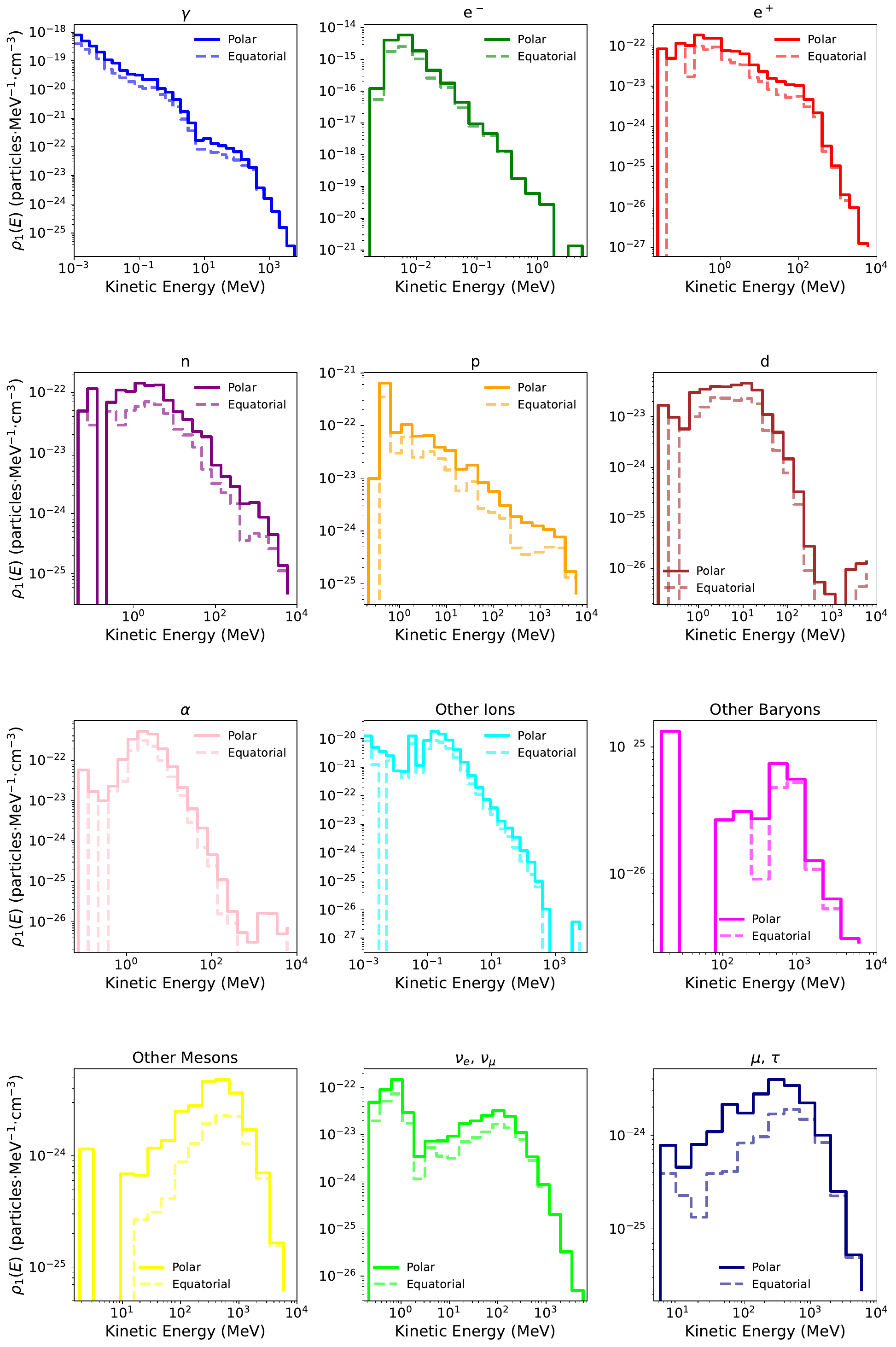}
 \caption{\small \justifying Volumetric production rate density $\rho_i(E)$ [particles~MeV$^{-1}$~cm$^{-3}$] as a function of kinetic energy for 12~particle species in Layer~1 (30--35~km). Solid lines represent the Polar scenario ($R_{\rm c} = 0.1$~GV) and dashed lines the Equatorial scenario ($R_{\rm c} = 15$~GV). All spectra are normalized per incident primary event. The Polar spectra systematically exceed Equatorial spectra by factors of $\sim$1.5--2, reflecting the 53.5\% Equatorial transmission probability. Electromagnetic particles ($\gamma$, $e^\pm$) dominate the production, followed by nucleons and leptons. The spectral shapes reveal characteristic electromagnetic shower cascades (power-law declines in $e^\pm$) and hadronic fragmentation (peaked neutron and proton spectra at $\sim$10--100~MeV).}
 \label{fig:volumetric_rate_layer1}
\end{figure*}

\begin{figure*}[htbp]
 \centering
  \includegraphics[width=0.80\textwidth]{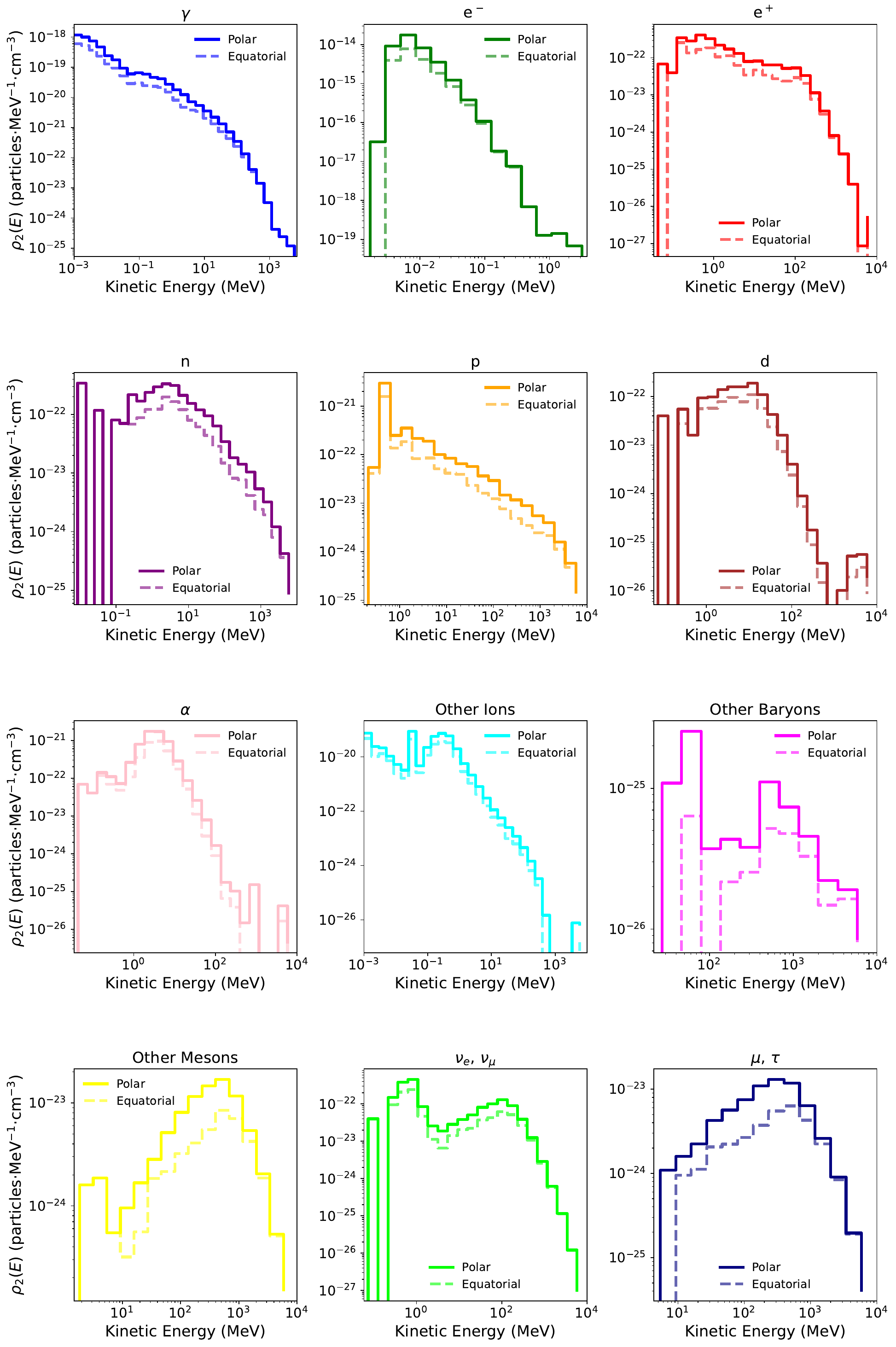}
 \caption{\small \justifying Volumetric production rate density for Layer~2 (25--30~km). Production rates increase by factors of 3--7 relative to Layer~1 as the cascade develops. The Equatorial-to-Polar ratio remains near 0.6 for most species, with the electromagnetic component beginning to show enhanced multiplication. Neutrino production becomes prominent, reflecting the development of hadronic decays ($\pi^\pm \to \mu^\pm + \nu_\mu$).}
 \label{fig:volumetric_rate_layer2}
\end{figure*}

\begin{figure*}[htbp]
 \centering
 \includegraphics[width=0.80\textwidth]{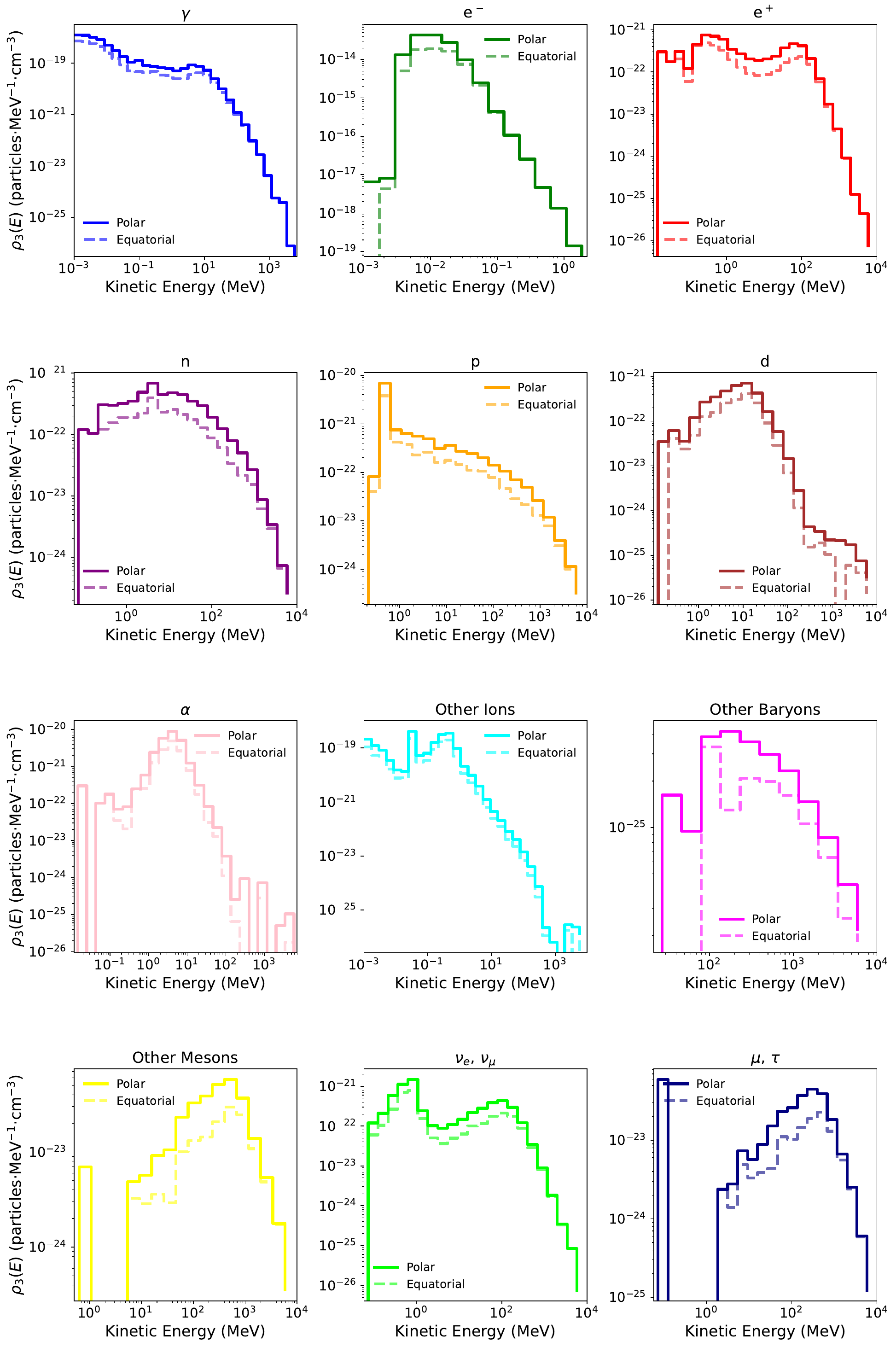}
 \caption{\small \justifying Volumetric production rate density for Layer~3 (20--25~km). Production rates increase dramatically, particularly for the electromagnetic component which grows by factors of 6--8 relative to Layer~2. The Equatorial-to-Polar ratio increases to 0.68 for total production, indicating enhanced cascade multiplication from harder Equatorial primaries. The photon spectrum extends to higher energies in the Equatorial case, reflecting deeper penetration of high-energy hadronic interactions that feed the electromagnetic cascade.}
 \label{fig:volumetric_rate_layer3}
\end{figure*}

\begin{figure*}[htbp]
 \centering
 \includegraphics[width=0.80\textwidth]{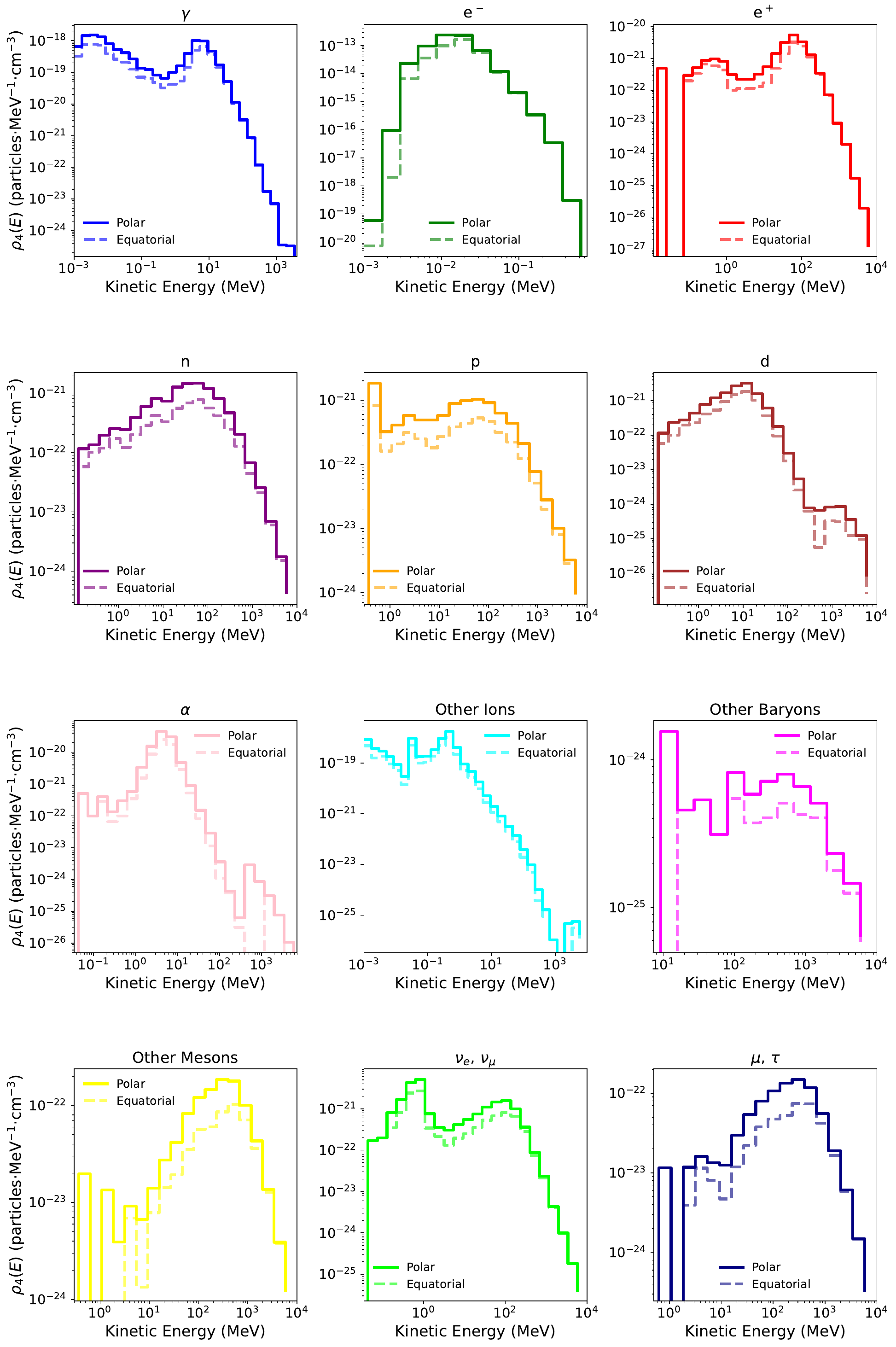}
 \caption{\small \justifying Volumetric production rate density for Layer~4 (15--20~km), the innermost layer near the Pfotzer maximum. Production rates reach their maxima, with the electromagnetic component achieving $\sim $680$\times$10$^{-20}$ (Polar) and $\sim $560$\times$10$^{-20}$ (Equatorial) particles~cm$^{-3}$ per incident primary. The Equatorial-to-Polar ratio increases to 0.82 for the EM component, the highest value observed across all layers and species. This demonstrates that spectral hardening from geomagnetic filtering drives substantially enhanced cascade development at depth, partially compensating for the reduced transmission probability. The convergence of Polar and Equatorial photon spectra at high energies ($>$100~MeV) indicates that the electromagnetic shower physics becomes increasingly independent of the primary spectrum once cascade multiplication dominates.}
 \label{fig:volumetric_rate_layer4}
\end{figure*}

This appendix presents the detailed layer-resolved volumetric production rate distributions, $\rho_i(E)$, for the four atmospheric shells (Figures \ref{fig:volumetric_rate_layer1}--\ref{fig:volumetric_rate_layer4}), illustrating the spectral differences between the Polar and Equatorial scenarios discussed in Section \ref{sec:volumetric_rate}.

\section{Supplementary Emerging Flux Spectra}
\label{app:emerging_flux_figures}

\begin{figure*}[htbp]
 \centering
 \includegraphics[width=0.95\textwidth]{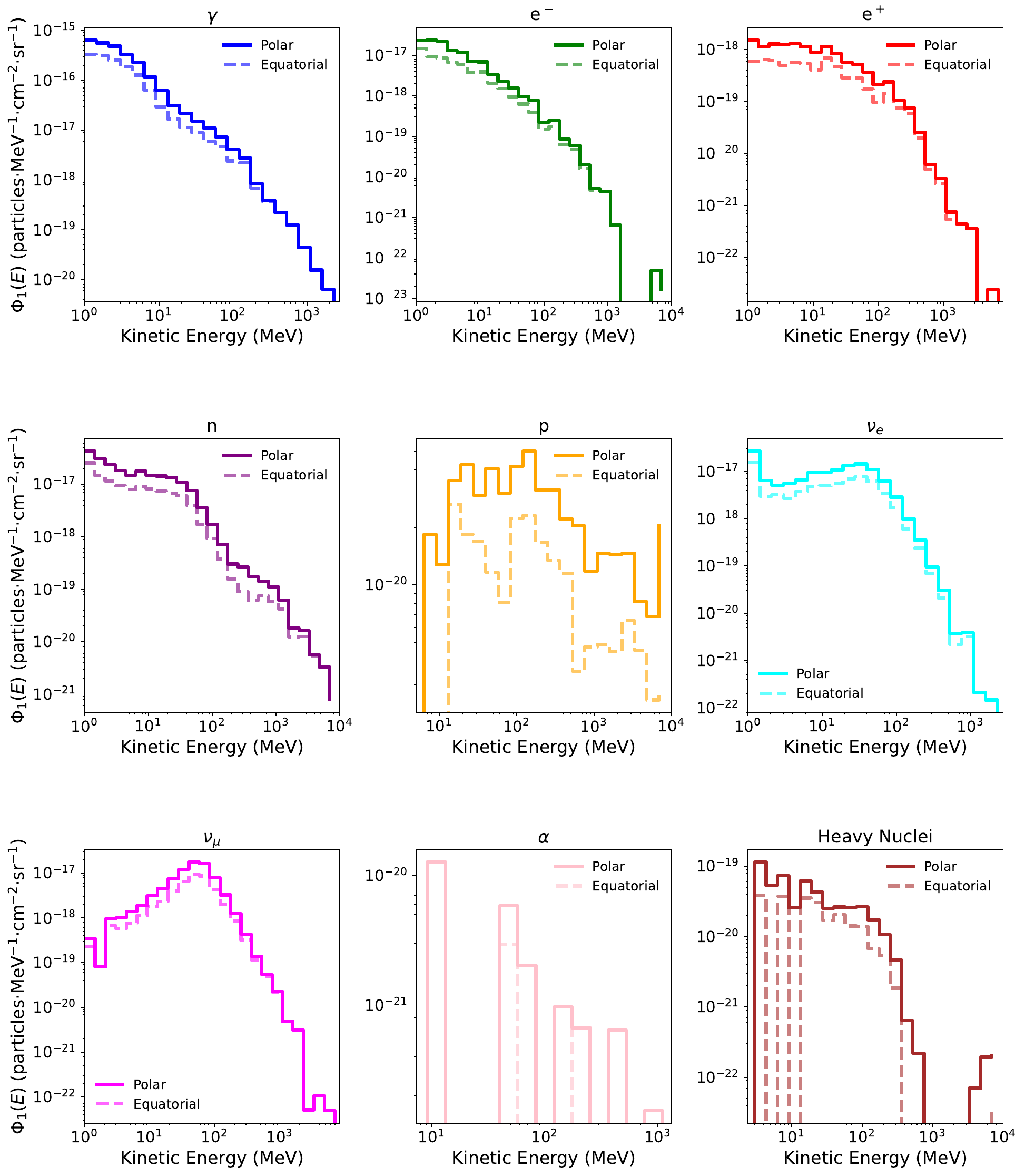}
 \caption{\small \justifying Differential emerging flux $\Phi_1(E)$ [particles~MeV$^{-1}$~cm$^{-2}$~sr$^{-1}$] as a function of kinetic energy for 9~particle species emerging from Layer~1 (30--35~km). Neutrinos ($\nu_e$, $\nu_\mu$) dominate the emerging flux, followed by photons and nucleons. The photon spectrum exhibits characteristic power-law decay, while neutrino spectra show peaked distributions near 10--100~MeV arising from charged pion decay chains. The Equatorial-to-Polar ratio of 0.64 for EM particles exceeds the overall ratio of 0.57, indicating modest spectral hardening effects even in the outermost layer.}
 \label{fig:emerging_flux_layer1}
\end{figure*}

\begin{figure*}[htbp]
 \centering
 \includegraphics[width=0.95\textwidth]{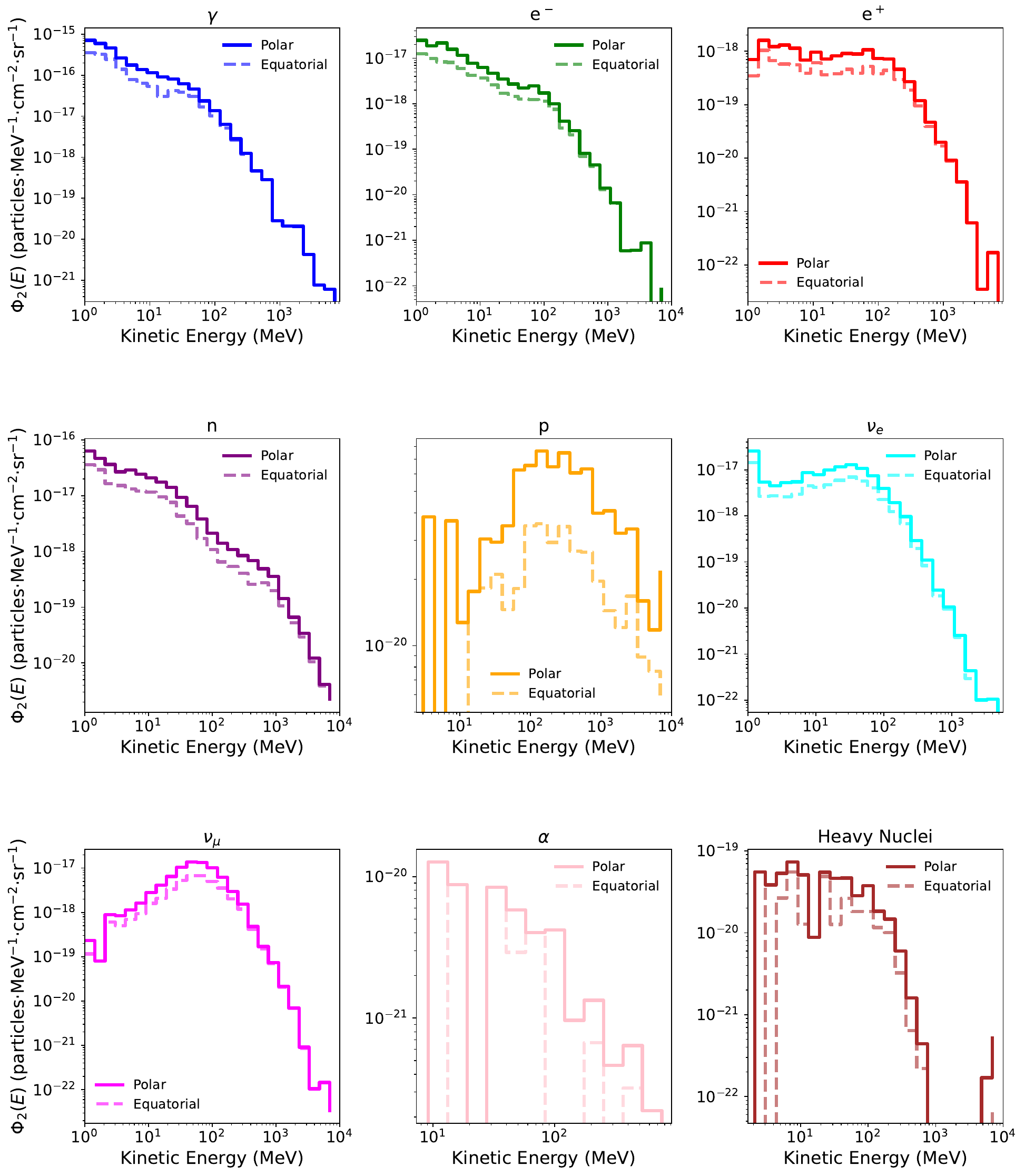}
 \caption{\small \justifying Differential emerging flux for Layer~2 (25--30~km). Fluxes increase by factors of 2--3 relative to Layer~1 as cascade multiplication intensifies. The Equatorial-to-Polar ratio improves to 0.60 for total flux and 0.65 for EM particles, demonstrating enhanced cascade development from harder Equatorial primaries. Photons dominate the EM component with characteristic $E^{-1}$ to $E^{-2}$ spectral slopes at high energies, reflecting bremsstrahlung and pair production processes. Neutrino spectra broaden and extend to higher energies, indicating contributions from higher-energy pion decays.}
 \label{fig:emerging_flux_layer2}
\end{figure*}

\begin{figure*}[htbp]
 \centering
 \includegraphics[width=0.95\textwidth]{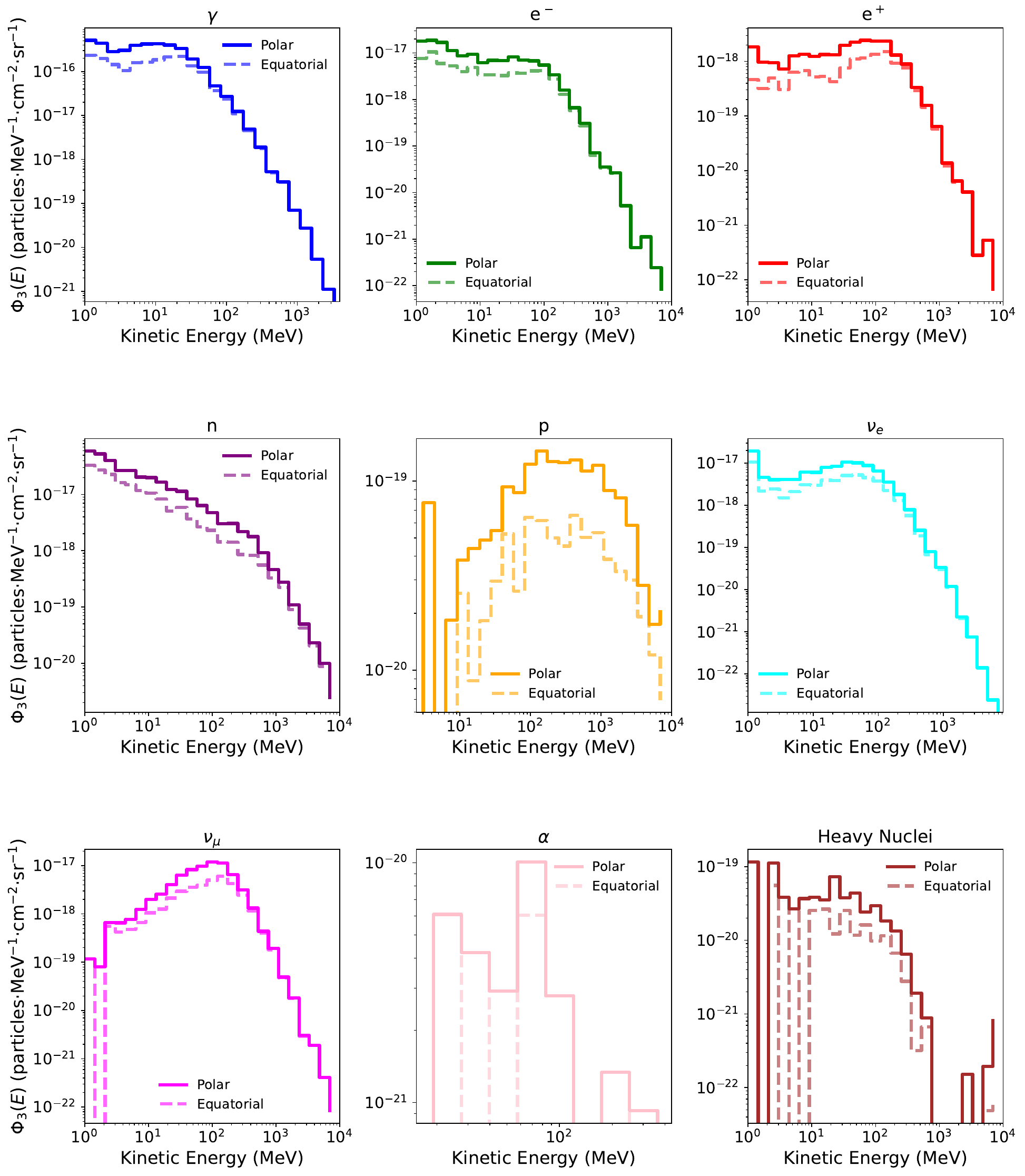}
 \caption{\small \justifying Differential emerging flux for Layer~3 (20--25~km), near the Pfotzer maximum region. Fluxes reach their maxima, with electromagnetic particles achieving the highest intensities observed across all layers. The Equatorial-to-Polar ratio increases to 0.67 for total flux and 0.72 for EM particles---the maximum enhancement observed in this study---demonstrating that spectral hardening most effectively amplifies cascade development at the altitude where atmospheric density optimizes shower multiplication. The photon spectrum extends prominently to high energies ($>$1~GeV) in the Equatorial case, reflecting deep penetration of energetic hadronic interactions. Neutrino spectra show enhanced high-energy tails, consistent with contributions from multi-GeV pion production.}
 \label{fig:emerging_flux_layer3}
\end{figure*}

\begin{figure*}[htbp]
 \centering
 \includegraphics[width=0.95\textwidth]{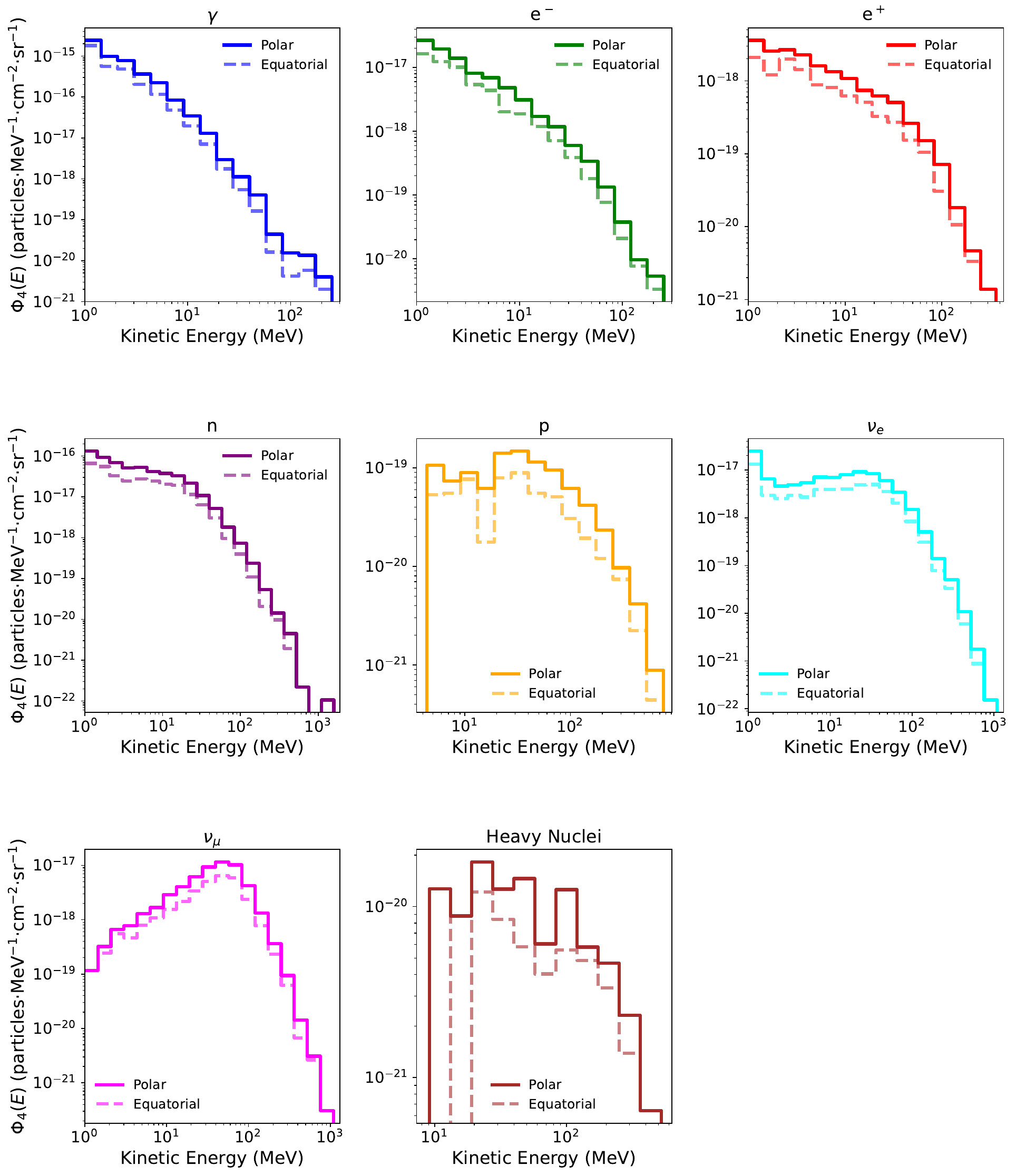}
 \caption{\small \justifying Differential emerging flux for Layer~4 (15--20~km), the innermost layer. Fluxes decrease dramatically by factors of $\sim$100 relative to Layer~3 due to atmospheric absorption. The Equatorial-to-Polar ratio returns to 0.57, closely matching the Layer~1 value, indicating that deep-penetrating fluxes are less sensitive to primary spectrum differences. Neutrinos dominate the composition (62\% of total flux in both scenarios), as weakly interacting neutrinos penetrate with minimal attenuation while EM and hadronic components are absorbed. The photon spectrum is substantially softer than in Layer~3, with the high-energy tail ($>$100~MeV) suppressed by more than two orders of magnitude. Neutron and proton spectra show characteristic evaporation peaks near 1--10~MeV, reflecting thermalized hadronic products from the overlying cascade.}
 \label{fig:emerging_flux_layer4}
\end{figure*}

This appendix provides the complete set of differential emerging flux distributions, $\Phi_i(E)$, for Layers 1 through 4 (Figures \ref{fig:emerging_flux_layer1}--\ref{fig:emerging_flux_layer4}). These spectra detail the energy-dependent transport of particles between atmospheric shells for both Polar and Equatorial scenarios.

\end{document}